\RequirePackage{fix-cm}
\documentclass[natbib,smallextended]{svjour3}       
\smartqed  
\usepackage{hyperref}
\usepackage{graphicx}
\usepackage{rotating}
\usepackage{amssymb,amsmath}

 \usepackage{aps-bibstyle}  
 \journalname{my journal}
\newcommand{\apj}{{Astrophys. J.}}

\begin{document}

\def\hst{{\sl HST~}}
\def\chan{{\sl Chandra~}}
\def\cxo{{\sl Chandra~}}
\def\ein{{\sl Einstein}}
\def\ros{{\sl ROSAT}}
\def\asc{{\sl ASCA}}
\def\Gam{$\Gamma$}
\def\si{$\simeq$}
\def\deg{$^\circ$}
\def\d300{$d_{300}$}
\def\micro{$\mu$}
\def\apj{ApJ}

\newcommand{\edot}{\dot{E}}
\newcommand{\pdot}{\dot{P}}
\newcommand{\lpwn}{L_{\rm pwn}}
\newcommand{\lpsr}{L_{\rm psr}}
\newcommand{\etapwn}{\eta_{\rm pwn}}
\newcommand{\etapsr}{\eta_{\rm psr}}
\def\lesssim{\mathrel{\hbox{\rlap{\hbox{\lower4pt\hbox{$\sim$}}}\hbox{$<$}}}}
\def\gtrsim{\mathrel{\hbox{\rlap{\hbox{\lower4pt\hbox{$\sim$}}}\hbox{$>$}}}}

\title{Pulsar-Wind Nebulae 
}
\subtitle{Recent Progress in Observations and Theory}


\author{Oleg Kargaltsev \and
        Beno\^it Cerutti \and
        Yuri Lyubarsky \and
        Edoardo Striani
}


\institute{O. Kargaltsev \at
              George Washington University, 105 Corcoran Hall, Washington DC, 20052, USA\\
              Tel.: +1 202-994-7225 \\
              Fax: +1 202-994-3001\\
              \email{kargaltsev@gwu.edu}        
           \and
           B. Cerutti \at
           Department of Astrophysical Sciences, Princeton University, Princeton, NJ 08544, USA\\
           \email{bcerutti@astro.princeton.edu}
            \and
           Y. Lyubarsky \at
          Ben-Gurion University, P.O.B. 653, Beer-Sheva 84105, Israel\\
           \email{lyub@bgu.ac.il}
           \and
           E. Striani \at
           Department of Physics, University of Torino, Via P. Giuria 1, 10125 Torino\\
           \email{edoardo.striani@iaps.inaf.it}
}

\date{Received: date / Accepted: date}

\maketitle

\begin{abstract}
In this review we describe recent observational and theoretical developments in our understanding of  pulsar winds and pulsar-wind nebulae (PWNe). We put special emphasis  on the results from observations of well-characterized PWNe of various types (e.g., torus-jet and bowshock-tail),  the most recent MHD modeling efforts, and the status  of the  flaring Crab PWN puzzle.
\end{abstract}

\section{Observations of PWNe.}

\subsection{Introduction}

Only $\sim 1\%$ of the total pulsar spin-down luminosity is emitted as pulsed electromagnetic radiation, the majority of the spin-down luminosity of a pulsar being carried away by a relativistic and highly magnetized pair plasma. These particles are generally believed to be accelerated and randomized in their pitch angle either upstream or at the pulsar wind termination shock. The radiation produced but these particles downstream of the termination shock is often seen as a pulsar-wind nebula (PWN).

Most of  the recent  progress in our understanding of PWNe has been spurred  by X-ray and TeV  $\gamma$-ray observations.  The majority of PWNe has been discovered in one of  these bands, and many are seen in both (see \citealt{2013arXiv1305.2552K}). To study PWN emission, it is important to disentangle the pulsar and pulsar wind contributions either by spatially  resolving the nebula from the pulsar or by isolating the PWN component in the spectrum (e.g.,  the PWN contribution is expected to dominate in TeV). High-resolution images from  {\sl Chandra X-ray Observatory} (\emph{Chandra} hereafter)  revealed two dominant PWN morphologies: torus-jet and bowshock--tail\footnote{Such classification is possible only for sufficiently bright and relatively nearby PWNe. }. In addition,  a few objects with varying and  puzzling morphologies can be seen in \chan images \citep{2008AIPC..983..171K}. The PWN properties (size, morphology, and spectrum) can be expected to depend on the pulsar parameters (spin-down properties, pulsar velocity, and the angles between the spin, magnetic dipole, and velocity vectors) and on the environment (e.g., ambient pressure, magnetic field, and radiation field). The limited angular resolution  of the ground-based TeV arrays (such as H.E.S.S. and VERITAS) does not allow us to detect  TeV emission from the same particles that produce bright and compact X-ray nebulae in the vicinity of the pulsar. Instead, TeV images reveal much larger structures filled with the aged  particles that may have accumulated over substantial part of pulsar's lifetime (see \citealt{2009ASSL..357..451D}). The TeV emission is usually attributed to  inverse Compton (IC) scattering of background optical/IR photons off aged electrons, although, in denser environments, the  contribution of neutral pion decay to the $\gamma$-rays emission could play an important role (if the pulsar winds indeed contain the so far elusive relativistic protons).

A general overview of PWN physics and X-ray observations was presented by \cite{2006ARAandA..44...17G} and \cite{2008AIPC..983..171K}, while PWN theory was recently reviewed by \cite{2014IJMPS..2860160A} and \cite{2014IJMPS..2860162B}.  Here we focus on some of  the most recent observational results\footnote{A more detailed overview of the observational properties of population of relic PWNe can be found in \cite{2013arXiv1305.2552K}.} and their implications   (\S1),  discuss the latest theoretical advances in MHD modeling (\S2), and review the non-MHD scenarios that can  explain the puzzling Crab PWN flares  (\S3).

\subsection{The Crab and Vela PWNe as prototypes of young PWNe in SNRs.}

The Crab and Vela PWNe are often considered to be archetypal representatives of torus-jet  PWNe. Since the Vela PWN is a factor of 10--20 older than the Crab PWN,  one can  look for evolutionary changes by comparing the two\footnote{ Since the PWN properties and evolution depend  on the environment, one should not forget that the progenitor and SNR properties may be quite different for the Crab and Vela pulsars. }.  PWNe of this type  are usually found around young pulsars whose velocities are smaller than the speed of sound  inside their host SNRs. These environments can  be characterized  by relatively high pressures and temperatures \citep{2010ApJ...709..507B}. There is also evidence that pulsars powering  torus-jet PWNe are likely to have substantial misalignment between the spin and magnetic dipole axes (e.g., Crab and Vela pulsars; \citealt{1999ApJ...522.1046M} and \citealt{2005MNRAS.364.1397J}, respectively)  which may play a pivotal role in  formation of this type of morphology.

\subsubsection{ Multiwavelength properties of the Crab}

\begin{figure}
\includegraphics[width=0.98\textwidth]{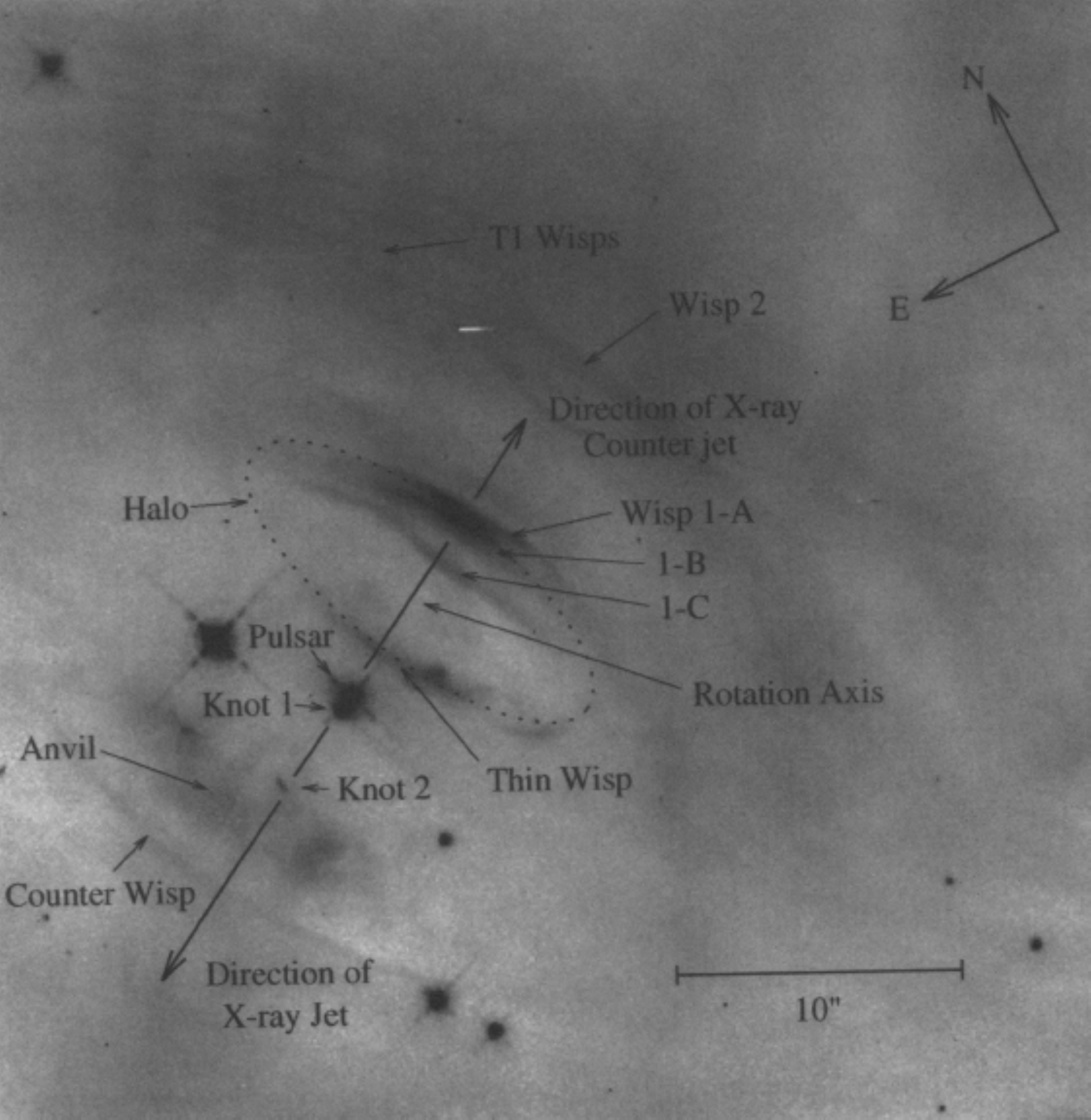}
\includegraphics[width=0.99\textwidth]{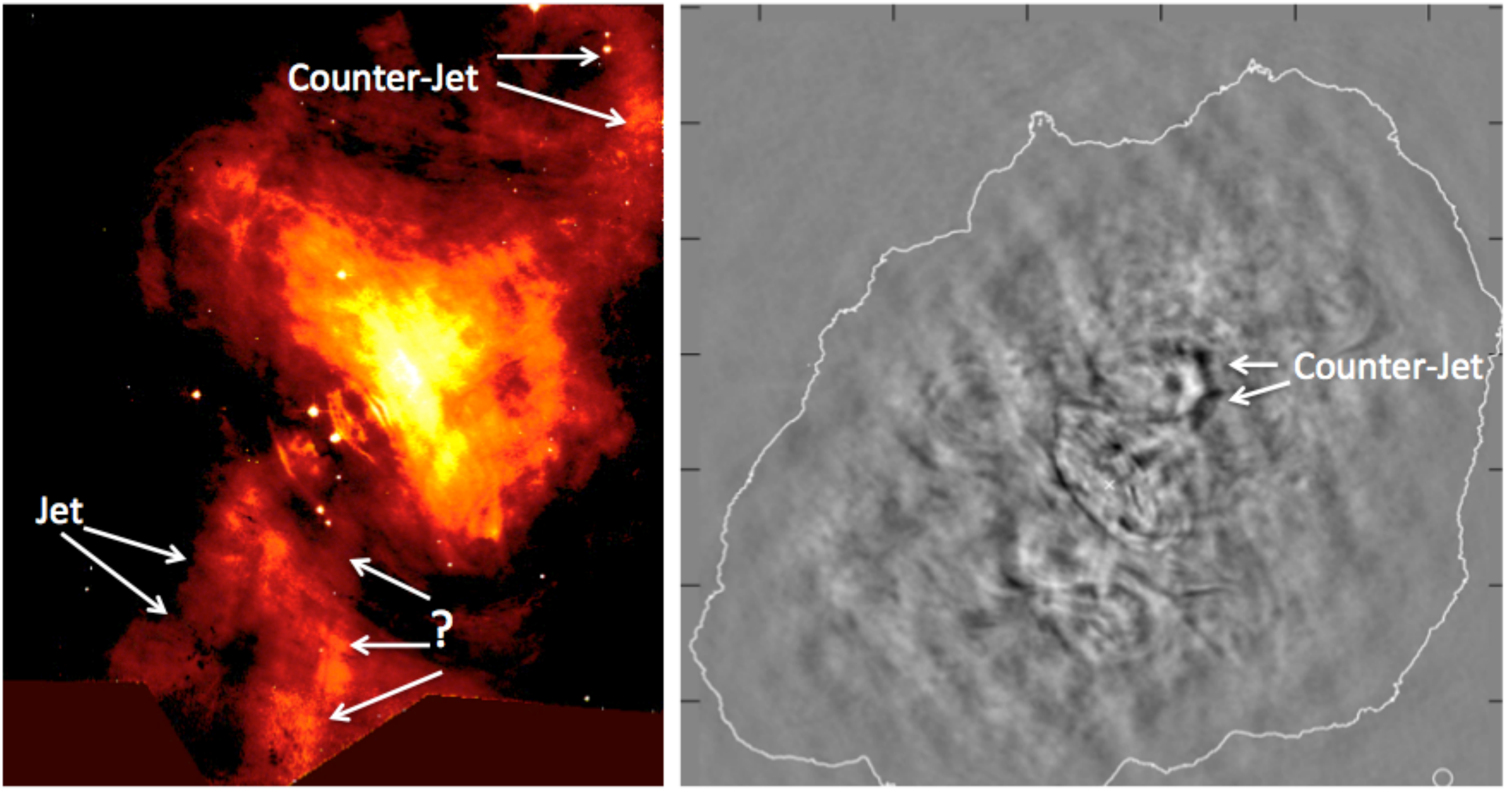}
\caption{The top panel introduces the most prominent features of the Crab PWN and their ``conventional'' names (from \citep{1995ApJ...448..240H}). The bottom left panel shows the 12.5 ks exposure image obtained with {\sl HST} ACS F550M. The image has been produced by a combining  series of auxiliary images obtained during the  09/2005--12/2005 polarimetry campaign \citep{2008ARAandA..46..127H}. The F550M filter avoids any strong emission lines and provides a relatively unobstructed view of the synchrotron nebula. The ``?'' mark enigmatic feature discussed in \S1.2.1 (also labeled as ``3'' in Fig.~\ref{crabmw}, top right panel). Notice that the feature labeled as the counter-jet in the F550M image also appears to show large variability in the difference image (shown in the bottom right panel) produced from 2 JVLA observations obtained on 2001 April 16 and 2012 August 26 (from \citealt{2014arXiv1409.7627B}). }
\label{linefree}
\end{figure}

The Crab nebula has been studied with nearly all major telescopes since its discovery (see \citealt{2008ARAandA..46..127H} and references therein). However, it was not until  the {\sl Hubble Space Telescope} ({\sl HST}; \citealt{1995ApJ...448..240H}) and \chan  \citep{2000ApJ...536L..81W} era when the intricate and complex structure of the nebula was revealed (see Figure \ref{linefree} for the feature nomenclature introduced by \citealt{1995ApJ...448..240H}). These observations have also shown that the bright inner part of the nebula is very dynamic, with apparent velocities corresponding to up to 0.5c (in projection onto the sky) as measured, e.g., from  the shifts in  wisp positions \citep{2002ApJ...577L..49H}.  The changes in the nebula are more complex than simple translational motion (e.g. steady expansion). They include variations in brightness (e.g., Inner Knot; \citealt{2005ApJ...633..931M}) and shape (e.g., Sprite; \citealt{2004ApJ...615..794B, 2002ApJ...577L..49H}). The wisp shapes can also be very different and  while most of the wisps can be described as a  ripple pattern with ripples moving away from the pulsar some of the wisps  appear at the same location (e.g., Thin Wisp in Figure  \ref{linefree}).   The prominent south-eastern (SE) jet (see Figure \ref{linefree})  also shows quite remarkable changes in its shape, based on X-ray images (taken over 14-year baseline), which could be explained by either precession of the curved jet or by the  motion of kinks  along the jet  \citep{2012int..workE...4W,2007ApJ...656.1038D}. Finally, the unexpected detection of $\gamma$-ray flares  by {\sl Fermi} LAT and {\sl AGILE} \citep{2012ApJ...749...26B,2011Sci...331..739A,2011Sci...331..736T} suggests a significant energy release rate (which can reach $4\times 10^{36}$ erg s$^{-1}$; \citealt{2012ApJ...749...26B})   on timescales of hours \citep{2013ApJ...775L..37M}; however, it was not possible so far to pinpoint the location of the flaring region  because of the lack of a ``smoking gun'' at lower frequencies. Consequently, the lack of information about the site of the flare in the PWN has led to a variety of models being suggested (see \S~\ref{sect_flares}). Given the lack of contemporaneous variability at lower frequencies,  it might be possible that some of the energy released during the process associated with the $\gamma$-ray flares  will manifest  itself as more gradual flux changes at lower frequencies\footnote{See animation at \href{http://home.gwu.edu/~kargaltsev/Crab.html}{http://home.gwu.edu/$\sim$kargaltsev/Crab.html}} occurring on much longer timescales (e.g., hard X-ray variability reported by \citealt{2013PASJ...65...74K, 2011ApJ...727L..40W}).

\begin{figure}
\includegraphics[width=0.999\textwidth]{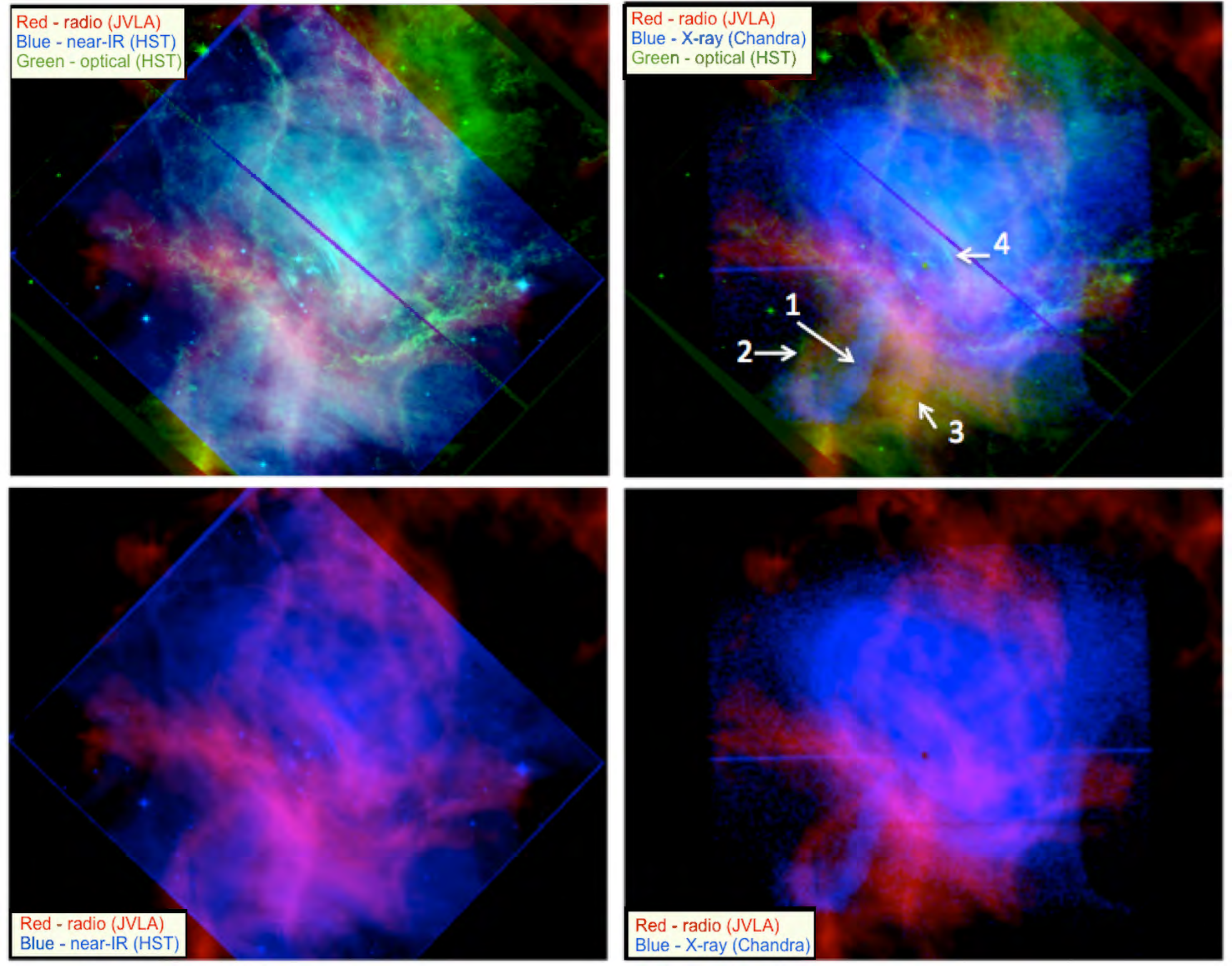}
\caption{False color MW images of the Crab PWN (see the legends in the panels) based on the observations described and analyzed in K+15. Numbers refer to the   PWN features mention in the text. }
\label{crabmw}
\end{figure}

\begin{figure}
\includegraphics[width=0.47\textwidth]{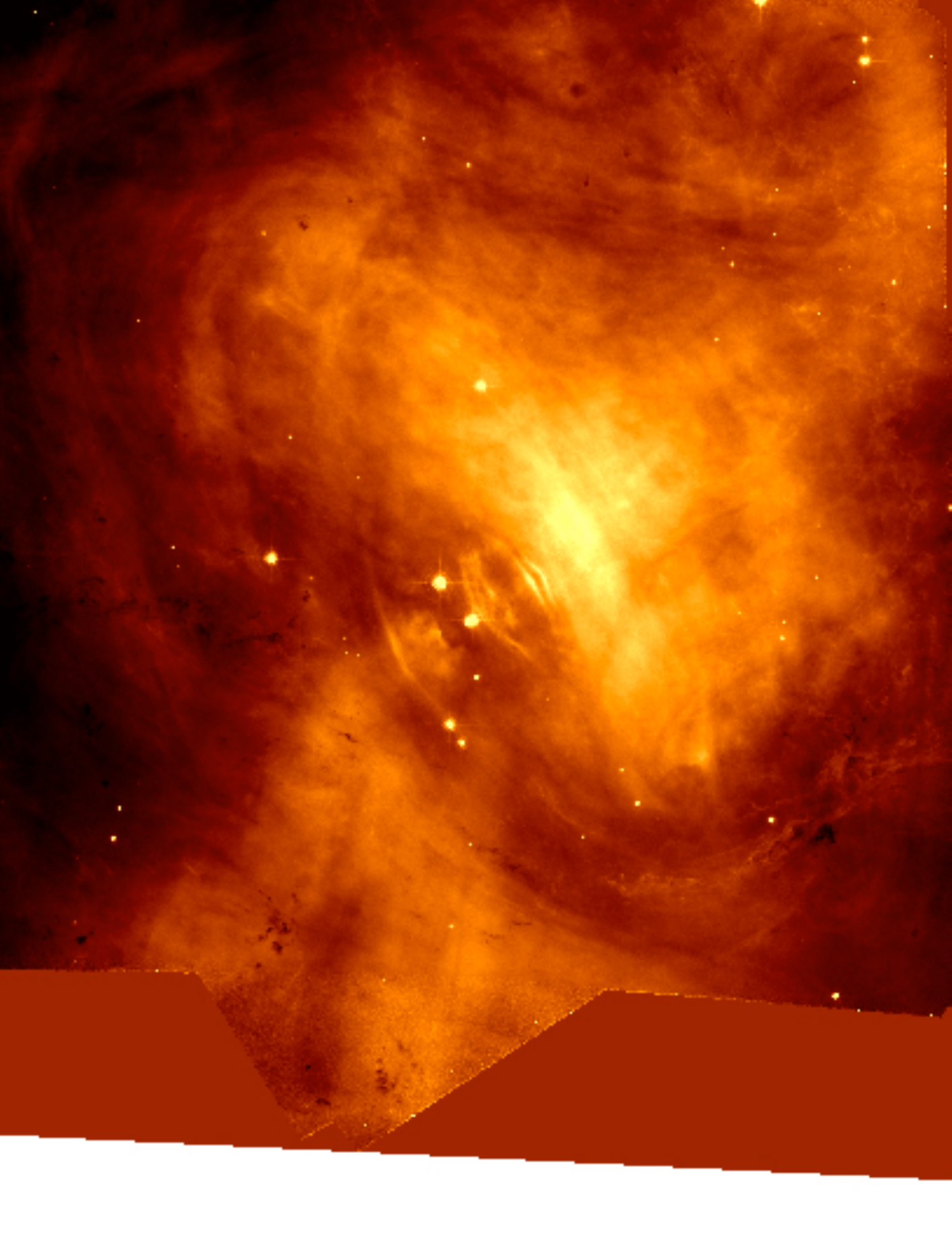}
\includegraphics[width=0.47\textwidth]{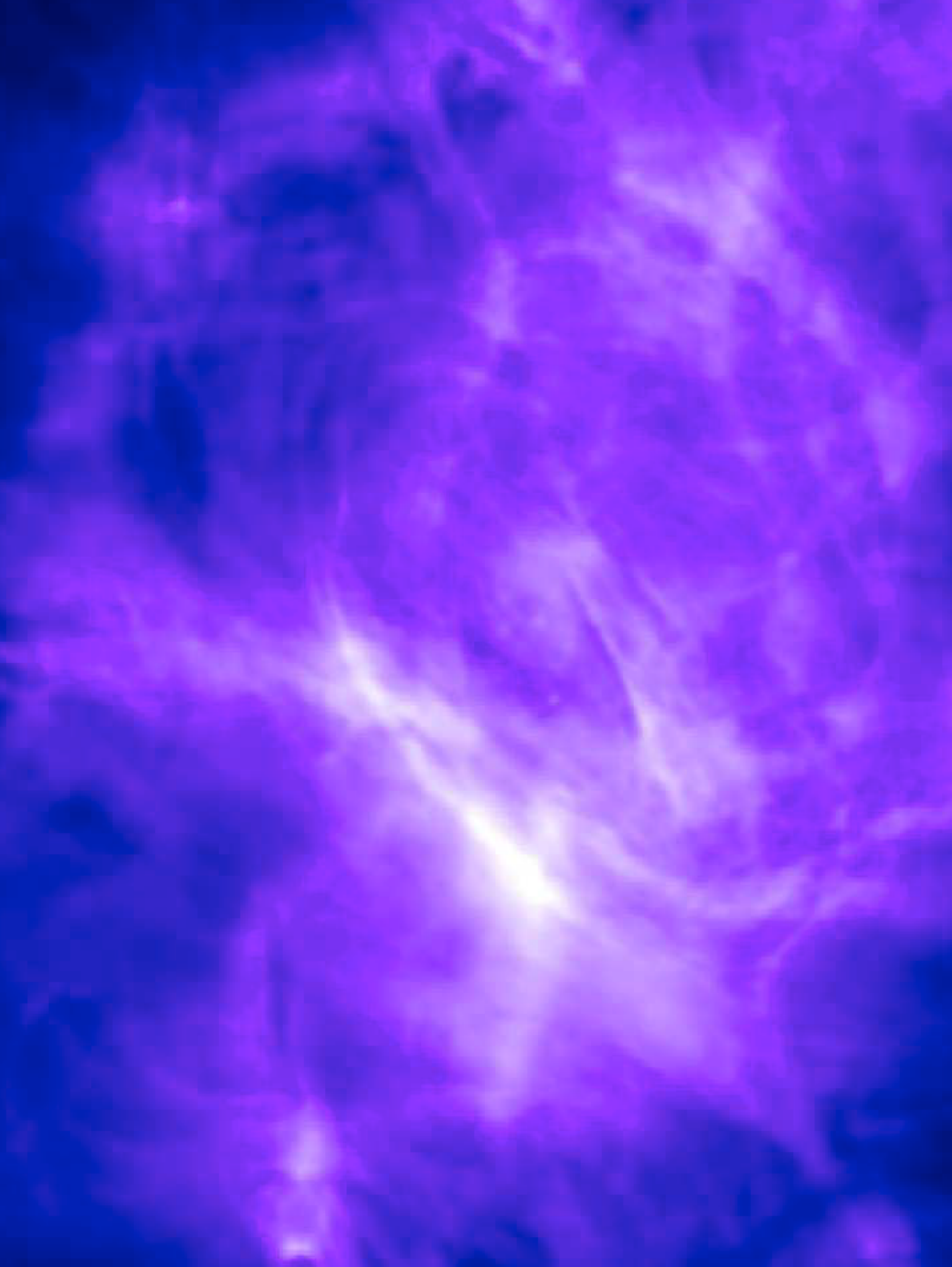}
\caption{ {\bf Left:} 12.5 ks exposure image of the Crab PWN obtained with {\sl HST} ACS F550M. The image has been produced by combining  a series of  auxiliary images obtained during the  09/2005--12/2005 polarimetry campaign \cite{2008ARAandA..46..127H}. The F550M   filter avoids any strong emission lines and provides a relatively unobstructed view of the synchrotron nebula. {\bf Right:}  JVLA image of the Crab (from  \citealt{2014arXiv1409.7627B}).}
\label{f55mvla}
\end{figure}

As a baseline for further comparison a multiwavelength (MW), high-resolution snap-shot of the  PWN  was obtained within a single day (Krassilchtchikov et al.\  2015 in preparation; K+15 hereafter).  Figure \ref{crabmw} shows {\sl Chandra}, {\sl HST} (NIR, optical), and Karl G. Jansky Very Large Array (JVLA)  images  from  this latest  MW campaign. These data  are ideally suited for measuring  the contemporaneous spectra of the prominent PWN  features.  Both the radio and  optical (broad-V band)  images reveal prominent filaments of which only some are coincident (which implies different MW spectra).  Also, the SE jet (``1'' in Fig.\ \ref{crabmw}), which is prominent in the X-ray image, appears to have counterparts in the optical and NIR images but not in the radio image, in agreement with the earlier findings of \cite{2004ApJ...615..794B}. On the other hand, a jet-like structure (``2'' in Fig.\ \ref{crabmw}) located to the east of the SE X-ray jet in the JVLA image does not have an X-ray or NIR counterpart and coincides with one of the thermal optical filaments\footnote{Overall, the large degree of correlation between the radio structure and optical filaments suggests  than most of the radio emission is related to the SNR filaments.}. Finally, in the NIR image, to the west of the bright X-ray jet (``1'' in Fig.\ \ref{crabmw})  there is another bright linear feature (``3'' in Fig.\ \ref{crabmw}) extending southward  from pulsar. The feature has a radio counterpart, a very faint X-ray counterpart, and does not coincide with any of the optical filaments. The faintness  of the X-ray counterpart 
  suggest that emission from this feature is a synchrotron continuum  may be produced by a cooled population of particles. It is, however, difficult to explain the presence of this low-energy feature in the conventional axisymmetric paradigm with equatorial and polar outflows, where only one jet and one counter-jet are expected\footnote{However, one can imagine that the jet  stayed at one position (``3'' in Fig.\ \ref{crabmw}) for a long time  and then relatively quickly moved  to the other position  (``1'' in Fig.\ \ref{crabmw})}.  On the other hand, the ``line-free''  (F550M) optical image shown in  Figure  \ref{linefree} supports the axial outflow paradigm by revealing better than ever  the other side of the axial ``backbone''  of the torus (dubbed as a counter jet, or NW jet).  We  note that at the  brightness/contrast  level chosen in   the F550M image (Fig.~\ref{linefree}, bottom left panel) the counterpart of the  SE X-ray jet is barely seen, while  the NW jet  and the SE ``jet-like'' feature (see above) are clearly seen.  Similar to the  SE X-ray jet, the  counter-jet appears to be hardly discernible  in the radio images (see e.g., right panel in Fig. \ref{f55mvla}) but it stands out in the 2-epoch difference image produced by  \cite{2014arXiv1409.7627B} and shown in the bottom right panel of Figure~\ref{linefree}. Finally,  the counter-jet is so faint in X-rays (if preset at all) that it cannot be discerned from the torus emission. Therefore, the frequency-dependent differences  in brightness between the jet and counter-jet appear to be more  complex than those expected for a simple scenario with the frequency-invariant Doppler boost (see e.g., \citealt{2013MNRAS.433.3325S}).  

K+15 found that the location of the bright optical/NIR wisp (``4''  in Fig.\ \ref{crabmw}) only approximately coincides  with the X-ray ring\footnote{The outer edge of the optical/NIR wisp is about $1''$ further away from the pulsar,  with the X-ray ring emission trailing behind or possibly being sandwiched between the bright wisp and the fainter wisp.}, and  the wisp brightness in the NIR image drops much faster with the distance from the symmetry axis of the nebula compared to the X-ray ring brightness.  This    
 supports   the \cite{2013MNRAS.433.3325S} findings  (based on earlier optical and X-ray monitoring) 
who concluded that  the X-ray and optical emission must be produced by different populations of particles.  Furthermore, according to  \cite{2013MNRAS.433.3325S}, the fits with the   Doppler-boosted tilted ring model require  noticeably different  (higher) flow velocities for the optical wisps ($\approx0.9c$) compared to the X-ray wisps, which made \cite{2013MNRAS.433.3325S}   question the simple ``boosted-ring'' model (see, however, \citealt{2015MNRAS.449.3149O}). K+15 also found that for most individual features of the PWN (e.g., wisps) the NIR-optical-FUV spectra are 
 harder than the contemporaneously measured X-ray spectra suggesting either an additional narrow spectral component or other kind of complex behavior between optical and soft X-rays.
 
The overall morphology of the Crab nebula has been reproduced in the relativistic magnetohydrodynamic  simulations with anisotropic energy flux (most recently by \citealt{2008A&A...485..337V, 2009MNRAS.400.1241C, 2014MNRAS.438..278P}). The simulated images display wisp-like features and other variable structures along the symmetry axis. In the MHD simulations these features are formed  due to the combination of flow dynamics (i.e. fluid vortices
produced near the TS due to magnetosonic oscillations; see e.g., \citep{2015MNRAS.449.3149O} and references therein)
and Doppler boosting.  The emission from the oblique termination shock that has been also associated with the Crab's Inner Knot (\citealt{2003MNRAS.344L..93K,2015arXiv150607282L, 2015arXiv150608121Y}; see, however  \citealt{2005ApJ...633..931M} and references therein for alternative interpretations). On the other hand,  some challenges to the MHD models still remain \citep{2014RPPh...77f6901B}.  The predicted  bright  arc  due to  the emission from the relativistic post-shock flow originating from the termination shock \citep{2003MNRAS.344L..93K,2006A&A...453..621D} is absent in the X-ray images where we instead see the patchy (likely consisting of multiple knots) inner ring which may or may not  appear to be brighter on the northwestern side (depending on the observation epoch). In this sense the optical (or NIR) images featuring a bright wisp NW of the pulsar appear to be in a better correspondence with the predictions of the MHD models. The models also predict co-spatial small-scale structures in the optical and X-rays while this is generally not observed (see above). Only parts of the X-ray inner ring  are seen in the optical and most (if not all) optical wisps also do not appear to have co-spatial X-ray counterparts (e.g., the Thin Wisp\footnote{Here we are following nomenclature introduced by \cite{1995ApJ...448..240H}, see Fig.~\ref{linefree} (top panel).} labeled in the top panel of Figure \ref{linefree} is lacking any nearby counterpart while other wisps are  offset from the possibly associated X-ray bright features; see \citealt{2013MNRAS.433.3325S} and K+15). Therefore, it yet remains to be shown whether more advanced  models 
  can fully capture  the rich  MW structure and variability of the Crab PWN. It seems that a complex injection spectrum may be required to achieve  this, hinting that there may be multiple acceleration sites throughout the PWN with possibly different acceleration mechanisms. The most recent, advanced 3D models predict somewhat disordered structure of the magnetic field and suggest the need for the in-situ  particle acceleration outside the termination shock region \citep{2014MNRAS.438..278P}. Diffusion transport may become more important in the case of disordered magnetic field. 

\subsubsection{Multiwavelength properties of the Vela PWN}

\begin{figure}
\includegraphics[width=0.999\textwidth]{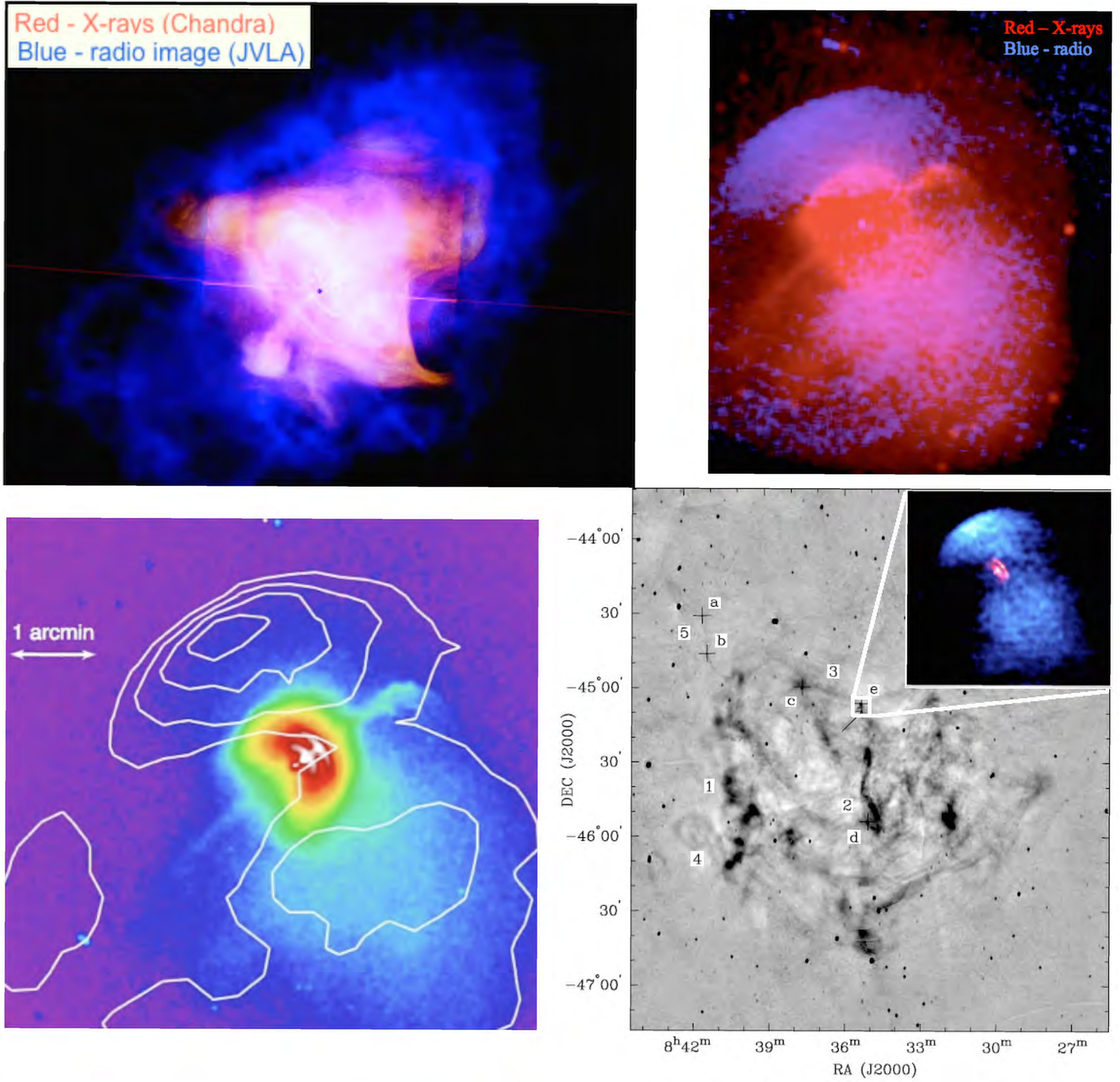}
\caption{Comparison of the Crab and Vela PWNe. The top panels show the combined X-ray (red) and radio (blue)  images of the Crab (top left) and Vela (top right) PWNe.  The bottom panels show X-ray  and radio  images of the Vela PWN. The {\sl Chandra} ACIS image (bottom left) shows that faint X-ray emission seems to fill in the radio lobes (shown by the contours).  The larger radio image on the right shows the entire Vela X complex (radio image from \citep{1997ApJ...475..224F}) within the Vela SNR. The inset shows the compact radio nebula (blue color) and brightest part of the X-ray nebula (red). }
\label{crab_vela}
\end{figure}

\begin{figure}
\includegraphics[width=0.999\textwidth]{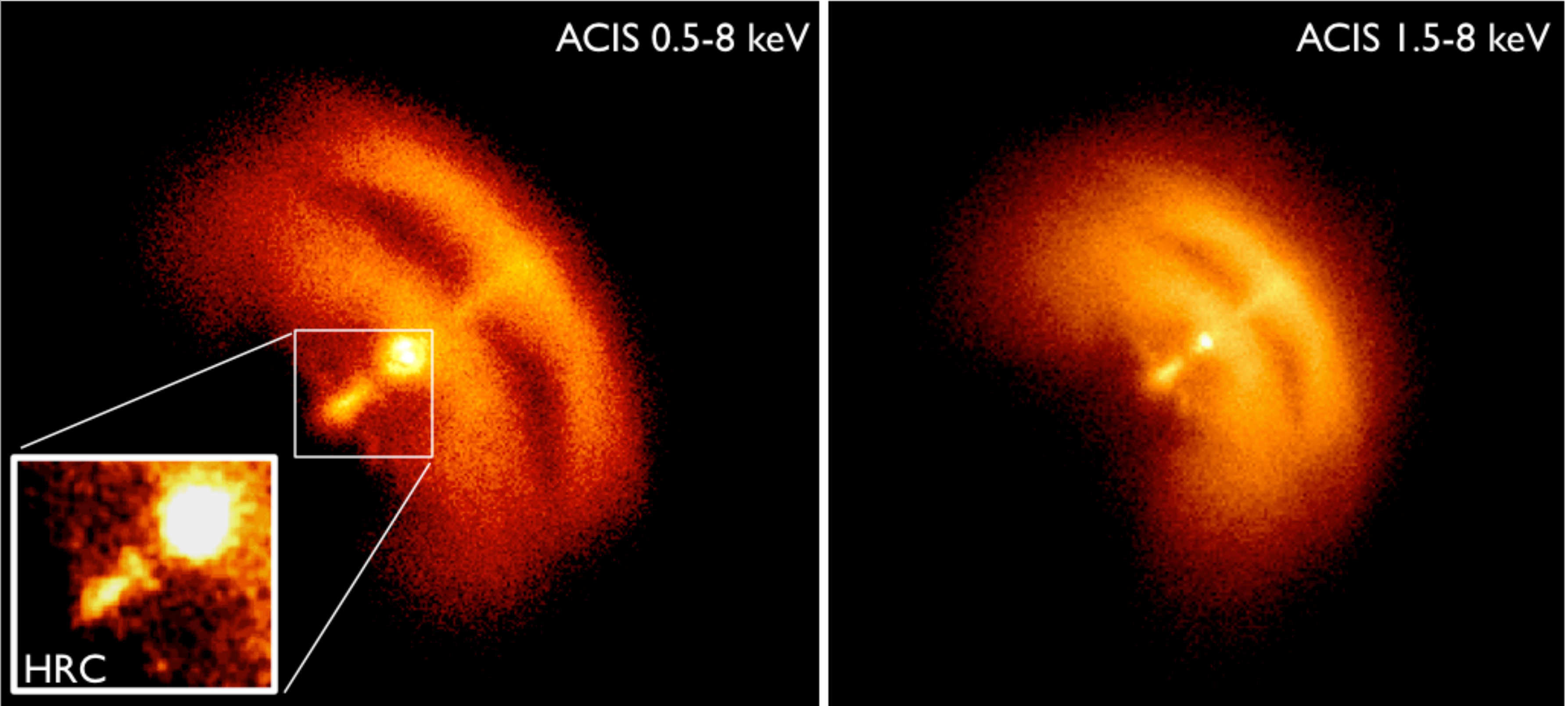}
\caption{ Deep {\sl Chandra} ACIS and HRC (inset) images  of the compact Vela PWN \citep{2013MmSAI..84..588L}.}
\label{veladeep}
\end{figure}

The Vela pulsar is a factor of 20 older than Crab, hence some evolutionary differences are expected. In addition,  the differences can be  attributed to a different progenitor type, different ISM, and different properties of the pulsar (e.g., magnetic field, angles between the spin and magnetic dipole axis, or the orientation and magnitude of the pulsar velocity). Figure 4 (bottom left) shows a deep {\sl Chandra} ACIS image of the Vela PWN (see also zoomed-in view of the compact nebula in Fig.~\ref{veladeep}) produced  by combining images from the latest observational campaign (comprised of  eight 40 ks observations taken with one-week intervals; see \citealt{2013ApJ...763...72D}. The overall morphology of the bright compact PWN can be described as an axisymmetric  double-arc structure with two axial jets having different brightnesses and widths. 
   The bright, compact X-ray PWN is located inside the larger double-lobe radio PWN \citep{2003MNRAS.343..116D}.  Interestingly, the radio lobes  appear to be filled  by  fainter X-ray emission which is particularly clearly seen  in the ACIS hard band (1-8 keV; see Figure 4) thus suggesting that the radiating particles have not cooled  much. Therefore, the pure advection model proposed by \cite{1984ApJ...283..694K} may need to be augmented with some other transport mechanism (e.g.,  diffusion) capable of moving  the energetic particles away from the pulsar more rapidly.  We also note that the bright, double-lobed radio nebula is surrounded by a much larger ($\sim 2^{\circ}$ in diameter) radio-emitting structure called Vela X (see \S1.5.4) which is filled with bright filaments and fainter diffuse continuum (Figure \ref{crab_vela}, bottom right).
   
\subsubsection{The Crab and Vela PWNe: similarities and differences}

The double-arc X-ray  morphology  of the Vela PWN is quite different from the Crab PWN, with its single ring, and torus (likely comprised of multiple wisps) seen in the ACIS images (see Fig.~\ref{crab_vela}). These differences can hardly be attributed to the larger age of the Vela PWN (because the X-ray emission from compact PWN mostly comes from freshly injected electrons) or different ambient (SNR) medium properties (these could become progressively more important further away from the termination shock).
It is also unlikely to be  due to the difference in the  angles between the pulsar spin axis and the line of sight because these angles are believed to be similar  ($127^{\circ}$ for Vela [\citealt{2001ApJ...556..380H}] and $\approx120^{\circ}$ for Crab [\citealt{2012ApJ...746...41W}]). The only other important parameters could be the angle between the spin and magnetic dipole\footnote{In principle,  the  magnetic field  may deviate from the dipolar configuration more for the Vela pulsar than for the  Crab pulsar. Indeed, braking indices, $n$, of the Crab  ($n=2.5$)  and Vela ($n=1.4$) are very different, and the Vela pulsar is much more ``glitchy'' compared to Crab.} axis (still rather  poorly constrained, e.g., $\sim43^{\circ}$ for Vela [\citealt{2005MNRAS.364.1397J}] and $\approx 45^{\circ}-70^{\circ}$ for Crab [\citealt{2013Sci...342..598L}])  and different pulsar velocities\footnote{The parameter that determines to what degree a PWN is affected by the pulsar motion is the ratio of the pulsar velocity to the local ambient sound speed (Mach number $\mathcal{M}$). The medium within the younger Crab SNR is hotter than in the Vela SNR.}  (the projected onto the sky velocities are $60d_{0.3}$ and $\sim120d_{2}$ km s$^{-1}$ for Vela and Crab, respectively\footnote{Here the distances are scaled as $d_{0.3}=d/(300$~pc) and   $d_{2}=d/2,000$~pc.}). These velocities are smaller than  the typical sound speed inside the young SNR, and therefore the Mach number $\mathcal{M}$ should be $\lesssim1 $ for both pulsars\footnote{It is likely that the Mach number is somewhat larger for the Vela PWN where we see some effect of the motion \citep{2003ApJ...591.1157P}.}.  Thus the differences in the compact PWN morphologies are more likely to be attributed to  the different   angles  between the spin and magnetic dipole axis.
 It is also possible that  the different degrees of deviation of the NS magnetic field  from that of  an ideal, centered dipole have  some impact on the PWN. Even for the compact parts of the PWNs some of the differences in X-ray morphologies could still be attributed to the longer synchrotron cooling time for the Vela PWN which is expected to have weaker magnetic field \citep{2003ApJ...591.1157P}.

We also note that the physical connection between the bright inner jets in the Vela PWN and the fainter large-scale outer jets (see e.g., Fig.~\ref{crab_vela} bottom left panel) has not been established yet.  The bright  SE inner jet  of the Vela pulsar suddenly appears out of the orthogonal bar-like feature (shock in the polar backflow?  \citealt{2003MNRAS.344L..93K}) at about  $5.4''$ from the pulsar and  then nearly as abruptly fades away at about $10.5''$ from the pulsar (see Fig.~\ref{veladeep}, left panel). In the very deep Vela PWN image the outer jets are  visible up to  much larger scales ($\simeq2-3'$); however, we do not see any smooth transition from the inner axial jets and hence we cannot establish a firm link between the two (except that both structures are extending along the PWN symmetry axis\footnote{In the \cite{2001AandA...379..551R} model the arcs are the traces of the particle beams from the two magnetic poles and the inner jets are  the Doppler-boosted projections of the beams.}). Note that simulated X-ray images based on MHD models (e.g. \citealt{2006A&A...453..621D,2009MNRAS.400.1241C}) do not show such structures. In the Crab PWN the dynamical feature called ``Sprite''  (possibly an analog of the bar feature in the Vela PWN) seems to be the place from which the SE jet originates. However, the Crab's SE jet does  not undergo any dramatic transitions in brightness until it  bends and terminates at about $1.2'$ from the pulsar.  Morphologically, this ``kinked'' jet resembles the outer jets of the Vela PWN rather than the straight and bright inner jets.  This leaves  no obvious analog to the bight inner jets of the Vela PWN in the Crab PWN (see, however, \citealt{2007ApJ...656.1038D}). As it has been discussed by  \cite{2003ApJ...591.1157P} and \cite{2013ApJ...763...72D}, the fact that the NW outer jet of the Vela PWN is brighter than the SE jet is at odds with the 3D orientation inferred from the arc brightness distribution.  The optical image of the Crab in Figure \ref{linefree} demonstrates a similar discrepancy assuming that the above mentioned  optical "counter-jet"  is the  actual NW jet.

Unfortunately, the compact Vela PWN has not been detected in the optical despite considerable efforts \citep{2003ApJ...594..419M,2014MNRAS.445..835M} hence direct comparison with the Crab PWN is not possible in this band. \cite{ 2013ApJ...775..101Z} recently reported  the extended feature seen in the NIR  (K$_s$ band)   which could be associated with bar at the base of the SE jet or  could be analog of the Crab's inner knot.  However, this result still requires confirmation.

It is also interesting to contrast the radio morphologies of the Crab and Vela PWNe. The filamentary structure of the Crab resembles that of Vela X, however, the latter  has a much larger angular extent ($r\approx60'$ for Vela vs.\  $\approx2.2'$) and it is much more asymmetric.   On the other hand, the Vela PWN is a factor of $7$ closer (hence, it should appear larger) and a factor of $10-20$   older (hence, it had more time to expand). Therefore, the different angular sizes are not surprising.  We also note that in the Vela PWN the TeV emission comes from the region of brightest radio filament \citep{1997ApJ...475..224F, 2006AandA...448L..43A} which is filled with the ejecta (based on the X-ray spectra; \citealt{2008ApJ...689L.121L}) thus providing denser target for the putative relativistic hadrons that might be present in the pulsar wind. Therefore, it may turn out that some of the prominent thermal Crab filaments are TeV bright. If confirmed, it could provide  evidence for the elusive hadronic component in pulsar winds \cite{1998MmSAI..69..989A}. Unfortunately, current resolution of the HESS and VERITAS telescopes does not allow to test this  hypothesis  (the existing data only suggest that the TeV emission is confined to within $<1.7'$ from the pulsar; \citealt{2006AandA...457..899A}).  Although it is plausible that the filamentary radio morphology from the Crab PWN is analogous to that of Vela X, there is no analogy in the Crab for the compact radio PWN found in Vela by  \cite{2003MNRAS.343..116D}.  The ATCA images reveal a double lobe structure, with the lobes being on each side of the X-ray PWN symmetry axis. The radio lobes, extending out to $3'-4'$, exceed  the size of the X-ray arcs  by a factor of 5  but  nonetheless they appear to be filled with the faint X-ray emission which is well seen in the harder (1$-$8 keV) band  (see Figure \ref{crab_vela}). Even if a similar  structure in the Crab would be smaller by a factor of $10-100$, it should have been resolved in the JVLA images (unless it is buried under the much brighter filamentary structure).  We also note that no wisp-like structures are seen in the X-ray, optical, or radio images of the Vela PWN.

\begin{figure}
\begin{center}
\includegraphics[width=0.5\textwidth]{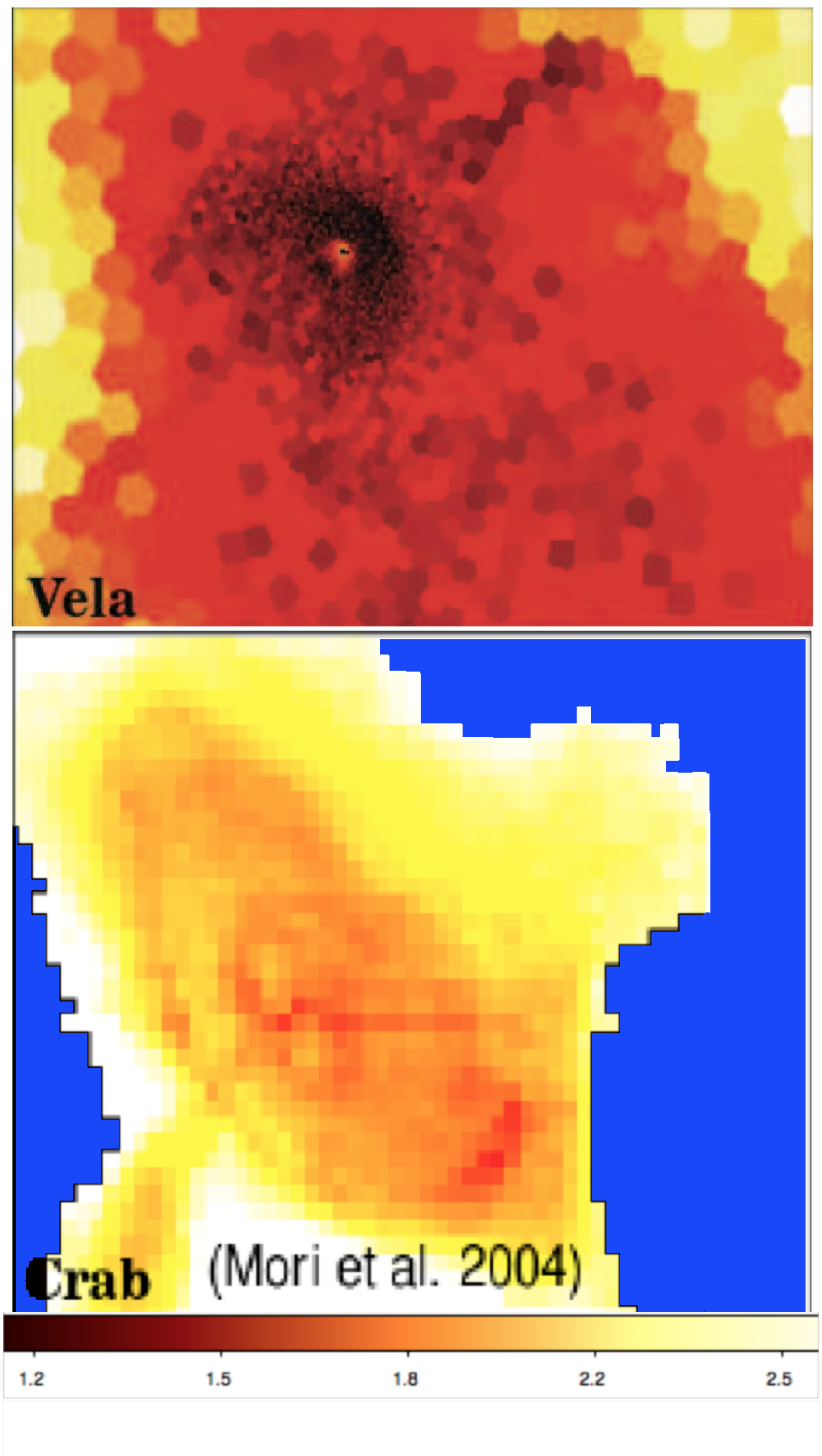}
\end{center}
\caption{ Photon index maps for Crab \citep{2004ApJ...609..186M} and Vela PWNe  \citep{2013ffep.confE...7K}  obtained from  {\sl Chandra} ACIS data. The color bar at the bottom shows photon index values. }
\label{pwnmaps}
\end{figure}

The X-ray spectral indices of the Crab and Vela PWNe are very different. Figure \ref{pwnmaps} shows spectral maps for the Crab and Vela PWNe. One can see that the spectra are the hardest (photon indices are the smallest) for the inner ring (in Crab PWN)  or for the arcs (in Vela PWN)  suggesting that these structures are associated with freshly injected accelerated particles. Interestingly, for the outer arc in Vela PWN the spectrum softens noticeably away from the symmetry axis while this is not the case for the inner arc in the Vela PWN or the inner ring in the Crab PWN.

\subsection{Bow shocks and tails: PWNe around supersonically moving pulsars.}

\begin{figure}
 \begin{center}
\includegraphics[width=0.999\textwidth]{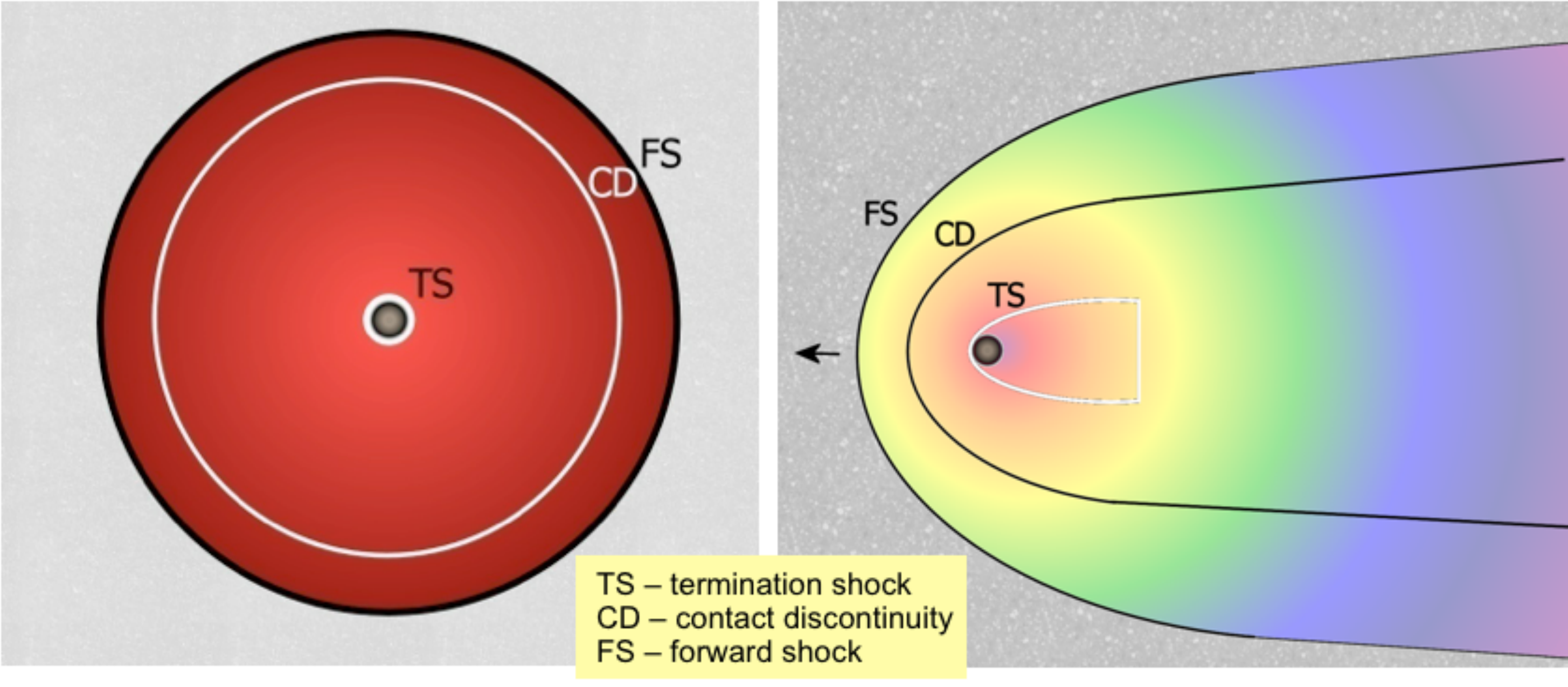}
\end{center}
\caption{ A schematic representation of  PWNe around the pulsar at rest (left) and supersonically moving pulsar (right) for an idealized case of isotropic pulsar wind.}
\label{schematic}
\end{figure}

Pulsar average 3D velocities have been found to be  $\sim400$ km s$^{-1}$ for isotropic velocity distribution (see \citealt{2005MNRAS.360..974H}). This implies that the majority of  pulsars  stay within their host SNR environment for  a few tens of thousands years although some particularly fast-moving pulsars can leave it earlier. Once the pulsar leaves the SNR\footnote{Alternatively, an old SNR can break-up and dissolve. },  it moves in a very different environment which has a much lower sound speed. For comparison, the sound speed in the middle-aged SNR(such as Vela SNR) can be on the order of a few hundred km s$^{-1}$.   The transition between the two very different environments should have a dramatic effect on the PWN of the high-speed pulsar. Once the pulsar is moving in the medium where its speed substantially exceeds the ambient sound speed (i.e.\  Mach number $\mathcal{M}=v_p/c_s\gg 1$, where $v_p$ and $c_s$ are the pulsar and sound speeds, respectively), the PWN shape should be strongly distorted by the ram pressure of the medium. If initially the wind was isotropic (this is obviously a great oversimplification, see above) the PWN would acquire  a cometary shape with the pulsar wind being confined to within the surface  formed by the contact discontinuity (CD) separating the shocked ambient medium and the shocked pulsar wind (see Fig. \ref{schematic}). Typically, it is assumed that the pulsar wind is  shocked in the termination shock (TS) at distances, $r_s$,  substantially smaller than the distance to the contact discontinuity, $r_{cd}$, even at the apex of the bowshock (see Figure \ref{schematic}). For  very fast moving pulsars  $r_{cd}$  may become so small that for some of the X-ray emitting  electrons the gyration radius, $r_g$, would become  $\sim r_s\sim r_{cd}$  which may lead to  leakage of the electrons from the apex of the bowshock (\citealt{2008AandA...490L...3B}; see also below). Numerical simulations by \cite{2005AandA...434..189B} indicate high flow speeds ($\gg v_p$) in the shocked pulsar wind outflow behind the moving pulsar, suggesting that an extended  pulsar wind tail should form\footnote{The simulations of \cite{2005AandA...434..189B} do not extend further than $25r_{cd,0}$ (where $r_{cd,0}$ is the scant-off distance at the apex of the bowshock) from the pulsar due to numerical challenges. Also, the model neglects the impact of the magnetic field on the flow dynamics. It is reasonable to expect that the pulsar tail physics  may have  some similarities with that of leptonic AGN jets for the case when the pulsar spin axis are parallel to its velocity vector (except that magnetic hoop stress may turn out to be larger in the case of pulsar tails). Therefore, some of the AGN jet simulations may be relevant for the pulsar tails. }. For realistically anisotropic pulsar winds (with equatorial and polar outflows), in addition to the Mach number, the appearance of the head of the bow shock PWNe and properties of the pulsar tails (to within a few $r_s$ from the pulsar) should also depend on the angle between the velocity vector and the spin axis of the pulsar.
  These effects have been investigated numerically by \cite{2007MNRAS.374..793V} in the limit of non-relativistic 3D hydrodynamics who found that  the bow shock morphology is only weakly affected by the pulsar wind momentum flux anisotropy but the morphology of the pulsar wind flow  in the tail is strongly affected\footnote{The simulations by \cite{2007MNRAS.374..793V} only extended for a few $r_{cd,0}$ from the pulsar, much smaller than the scales of extended tails seen in X-rays).}. On the other hand, the ambient medium non-uniformity was found to be greatly affecting the bow shock symmetry and shape. Overall, \cite{2007MNRAS.374..793V} concluded that ``the anisotropy of the wind momentum flux alone cannot explain the observed bow shock morphologies''. The simulations also show that Kelvin-Helmholtz (KH) instabilities can develop if the ambient medium exhibits a large pressure gradient. These could be further amplified  if the  relativistic nature of the pulsar wind flow is taken into account due to the increased velocity shear \citep{2005AandA...434..189B}. In such situation it is possible that the shocked ambient material can be entrained in the pulsar wind flow altering its structure, dynamics, and emission properties. The entrainment of ambient matter in the pulsar wind is largely an unexplored area (see, however, \citealt{2003MNRAS.339..623L,2015arXiv150501712M}). Further simulations of bow shock PWNe combining these effects (relativistic velocities, ambient medium non-uniformity and entrainment, pulsar wind anisotropy and dynamical role of the magnetic fields, 3D geometry and instabilities) can provide a realistic picture for comparison with the observations.  It may also be possible to make progress by advancing the analytical models of these outflows.  \cite{2005ApJ...630.1020R}  has  constructed a model of a pulsar magnetotail for the axisymmetric case  (the pulsar velocity is co-aligned with its spin axis). In this model the pulsar  wind remains collimated at large distances from the pulsar forming a magnetotail  where an equipartition is reached between the magnetic energy and the relativistic particle energy. The model predictions for the shape of the magnetotail  appear to  agree with  the data in some cases (e.g., PSR J1101--6101; \citealt{2014arXiv1410.2332H} and PSR J1747--2958; \citealt{2004ApJ...616..383G}, \citealt{2005AdSpR..35.1129Y}) and disagree in  others (e.g., PSR J1509--5850; \citealt{2008ApJ...684..542K}), possibly, discriminating between the  axisymmetric and non-axisymmetric cases.

\begin{figure}
\includegraphics[width=0.999\textwidth]{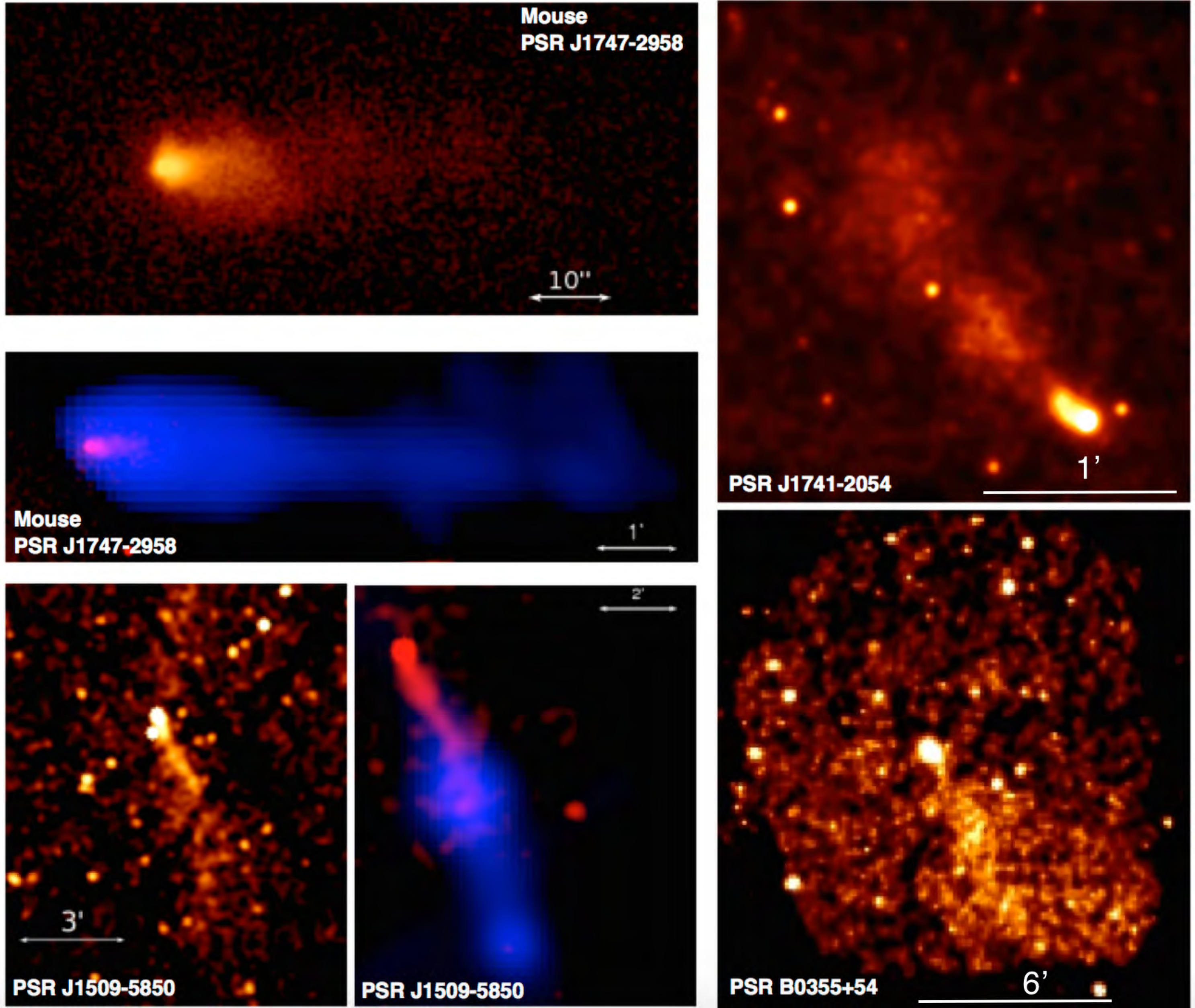}
\caption{ X-ray (red) and radio (blue) images of pulsar tails produced from archival \chan and JVLA data. The J1509-5850 radio image is based on the  \citep{2010ApJ...712..596N} analysis of Australia Telescope Compact Array observation. }
\label{tailsheads}
\end{figure}

\begin{figure}
\includegraphics[width=0.66\textwidth]{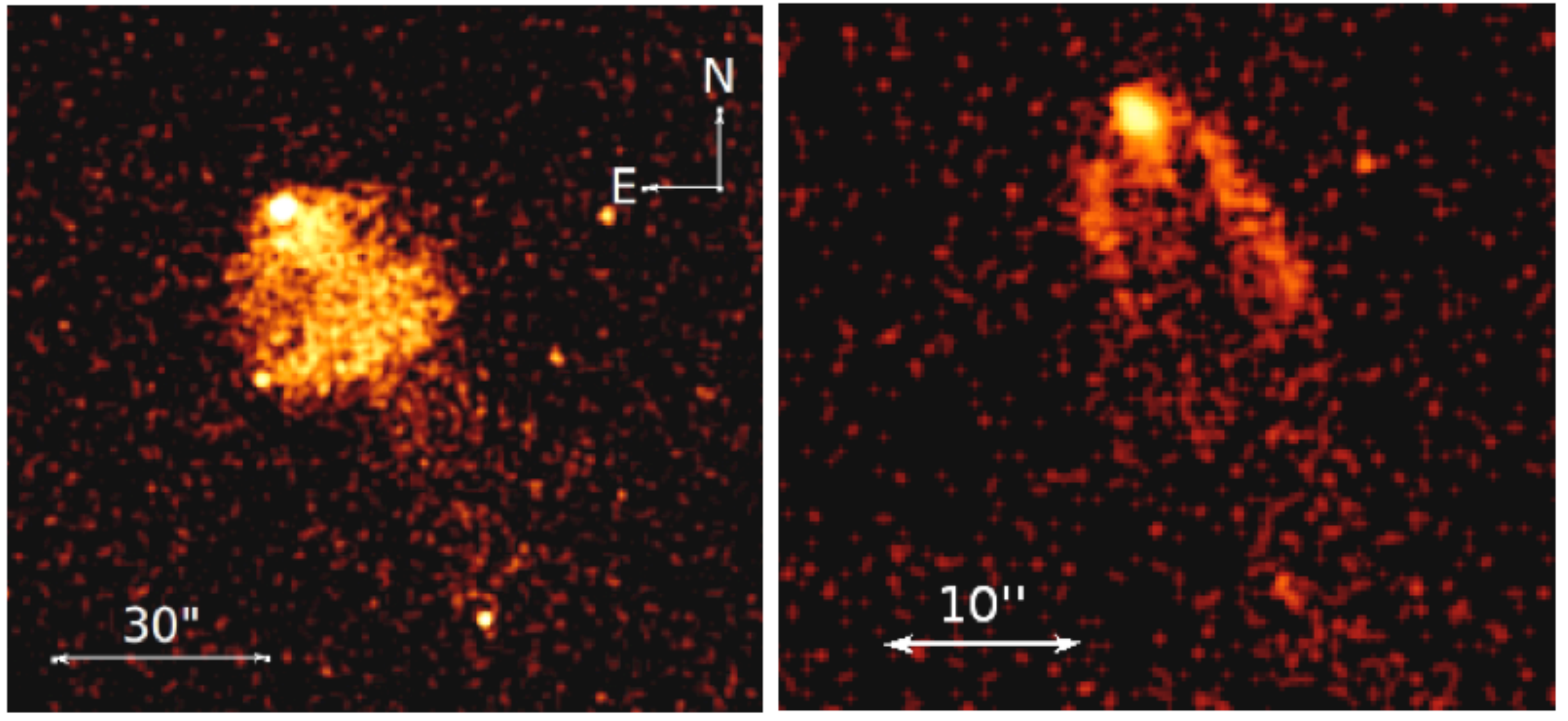}
\includegraphics[width=0.33\textwidth]{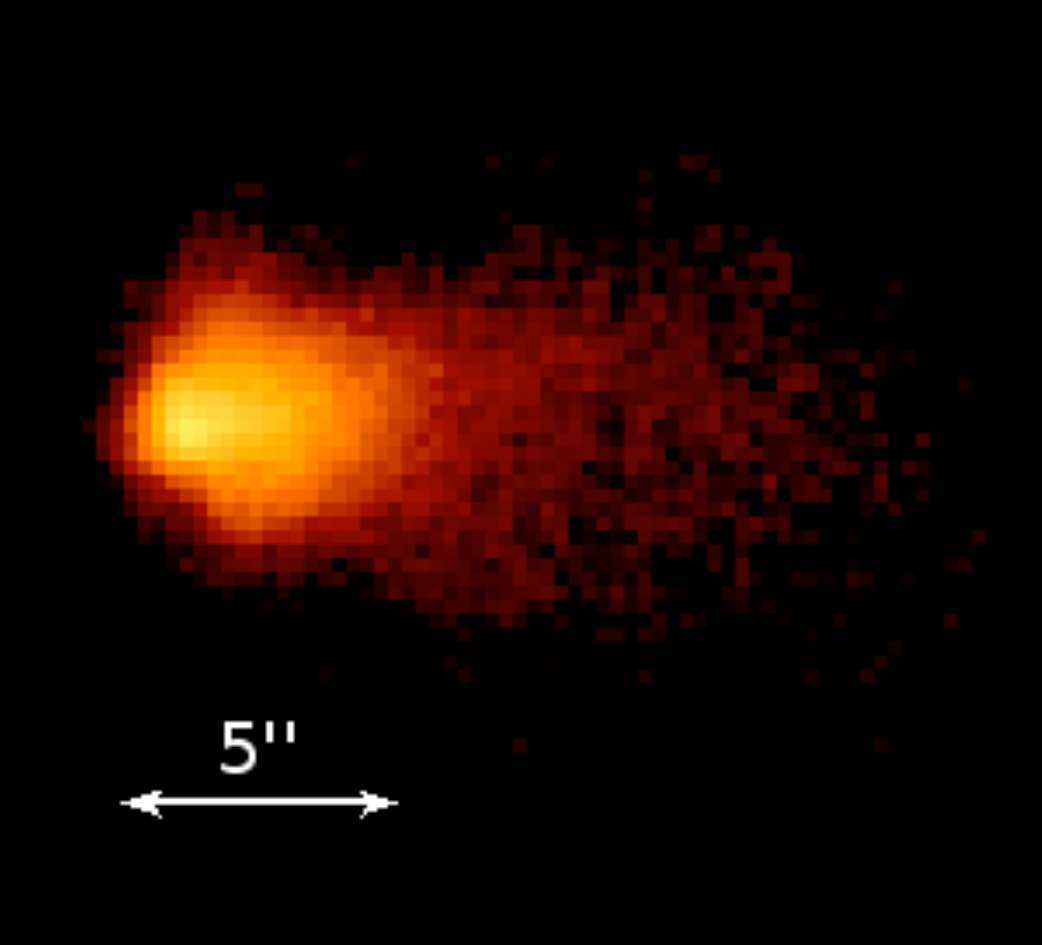}
\caption{ {\sl Chandra}  ACIS  images of the head regions of B0355+54, J1509--5058,  and Mouse  PWNe (left to right). Notice very  different morphologies of  B0355+54 and J1509--5058 PWNe. The images are produced from archival \chan  data.}
\label{heads}
\end{figure}

\begin{figure}
\includegraphics[width=0.999\textwidth]{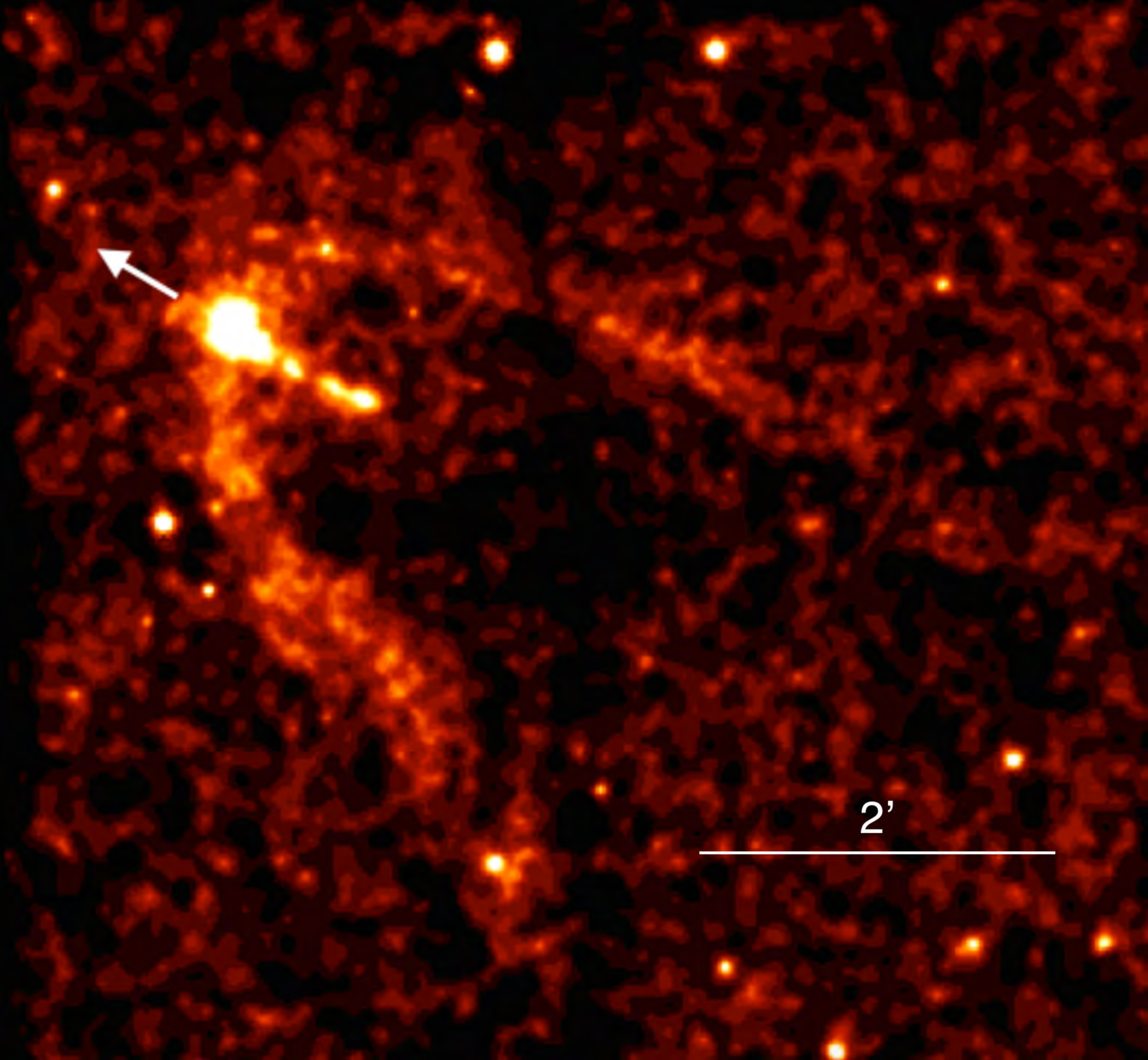}
\caption{ 570-ks \cxo ACIS image of  Geminga PWN (Posselt et al.\ in prep.). }
\label{geminga}
\end{figure}

\begin{figure}
\includegraphics[width=0.999\textwidth]{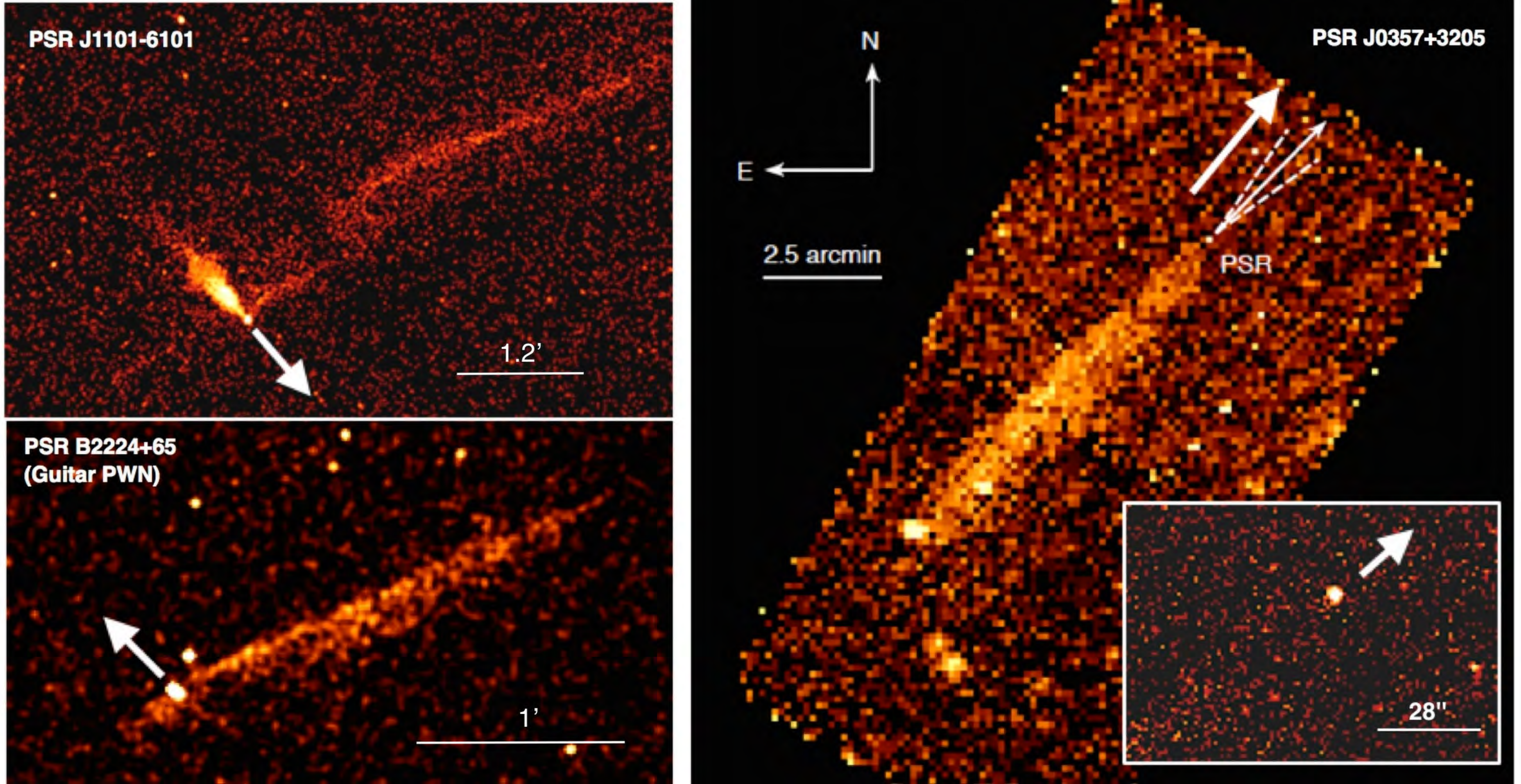}
\caption{ Puzzling PWN morphologies are seen  in  the \chan images of three high-speed pulsars. The arrows show pulsar proper motion directions. In two cases (left panels) the extended structures are orthogonal to the pulsar proper motion. For PSR J0357+3205 (right panel) the extended X-ray emission brightens further away from the pulsar while it is very dim in the immediate vicinity  of the pulsar (shown in the inset). See \cite{2014AandA...562A.122P,2013ApJ...765L..19D,2010MNRAS.408.1216J} for detailed analysis.}
\label{puzzling}
\end{figure}

{\sl Chandra} and {\sl XMM-Newton} observations provided for the first time  X-ray images of PWNe around supersonically moving pulsars (see examples in Figures \ref{geminga}, \ref{tailsheads}, \ref{heads}, and \ref{puzzling}).  Some of these images display structures (see e.g.,  Figs.\ \ref{tailsheads} and \ref{heads} ) that are broadly consistent with the theoretical expectations. Indeed,  in the X-ray  images shown in Fig.\ \ref{tailsheads} one can identify  bright PWN heads accompanied by  much fainter extended tails (see Fig.\ \ref{tailsheads}).  Interestingly, the X-ray bright PWN head may or may not be bright in the radio [c.f.\  PWN of PSR J1747-2958 (``Mouse'') and PWN of PSR J1509--5058]. Radio polarimetry  of  two extended tails  (Mouse and  J1509--5058; \citealt{2005AdSpR..35.1129Y} and \citealt{2010ApJ...712..596N}) shows that the magnetic field direction is predominantly transverse in the case of the J1509--5058 tail and  aligned with the tail in the case of  the Mouse tail. This may suggest that the spin axis is more aligned with the velocity vector for J1509--5058 than  for the Mouse pulsar (see Fig.~3 from \citealt{2005ApJ...630.1020R}).  Recently, deep, high-resolution observations with {\sl Chandra}  revealed interesting structures of the bright  bow shock PWN heads. Images of  J1509--5058 and B0355+54 PWNe, shown in Figure \ref{heads}, exhibit contrasting morphologies.  The head of the B0355+54 PWN shows symmetric, filled ``mushroom cap'' morphology with  emission being somewhat brighter near the center than on the sides. On the other hand, the head of the  J1509--5058 PWN,  comprised of two bent arcs resembling a bow structure (the arcs, however, do not quite connect to the pulsar),  is mostly ``empty''  except for the slight extension just behind the pulsar. This structure remarkably resembles that of the nearby ($d\approx250$ pc) Geminga PWN (see Figure \ref{geminga}) as it would be seen at a much larger distance of J1509--5058 ($d\approx 4$ kpc).  The bow-shaped X-ray emission can either be associated with the forward shock in the ambient medium (unlikely, because the pulsar velocity must be very high to produce X-rays by heating  ISM) or   pulsar jets. In the latter scenario the outflows from  J1509--5058 and Geminga must be dominated by the luminous jets rather than the equatorial component (cf.\ Crab and Vela PWNe). This may be difficult to reconcile in the  \cite{2004MNRAS.349..779K} model where the jet formation is intimately connected to the diverted by the magnetic field hoops stress equatorial outflow (backflow) which helps to collimate  the polar outflow.  Furthermore, the  recent 3D simulations (see \S2.2 and 2.3) suggest reduced axial compression and weaker jets compared to the 2D simulations. On the other hand, most numerical simulations are designed to reproduce the Crab and Vela structures with a large angle between the NS magnetic dipole and spin axis.  If this angle is small, the outflow dynamics could be substantially different. If the side arcs of J1509 are indeed  jets, it would also be difficult to explain the ordered helical magnetic field morphology in the extended  tail (revealed  by the radio polarimetry; \citealt{2010ApJ...712..596N}) because such a structure would be more natural for the axially symmetric case  \citep{2005ApJ...630.1020R}. Thus, although it is plausible that qualitative morphological differences  in the appearances of compact PWNe can be attributed to the geometrical factors (i.e.\ angles between the line-of-sight, velocity vector, spin axis, and dipole axis), these dependencies are yet to be understood.

Particularly interesting and puzzling is the transition region between the bright PWN head and the faint extended tail. For instance, in the B0355+54 PWN (``Mushroom'') the drop in the surface brightness at the trailing edge of the Mushroom ``cap'' is nearly as sharp as at the leading edge (this makes it unlikely to be due to the synchrotron burn-off).  In the conventional (isotropic pulsar wind) bow-shock tail models this could be associated with the back surface of the bullet-shaped termination shock \citep{2005AandA...434..189B, 2004ApJ...616..383G}. However, this interpretation does not appear to work for the  J1509--5058 PWN which lacks the emission from putative  back surface of the termination shock. The transition is also much smoother in the Mouse PWN (see Fig.\  \ref{heads}, right panel). The B0355+54 PWN image also shows much fainter, narrow ``stem''  attached to the Mushroom cap which makes it tempting to associate the bright trailing edge of the Mushroom cap with the  equatorial termination shock that has been pushed back by the ram pressure. In this scenario the stem and the brighter middle part of the Mushroom cap would be associated with polar outflow (a jet).  
 However, even in this case the  drop in the brightness at the trailing edge of the Mushroom cap may be too abrupt. For instance, in the deep \cxo images of the compact Vela PWN one can see the effect of the motion onto the inner ring (commonly associated with the termination shock) with the particles being blown back off the inner ring (see Fig.\ \ref{veladeep}, right panel) . We do not observe such a  smooth transition behind the Mushroom ``cap'' (Klingler in prep.).

The PWNe behind several very fast-moving pulsars display puzzling morphologies (see Figure \ref{puzzling}). These are  the ``Lighthouse nebula''  with PSR J1101-6101  \citep{2014arXiv1410.2332H, 2011AandA...533A..74P, 2012ApJ...750L..39T},  the ``Guitar nebula'' with PSR B2224+65 \citep{2010MNRAS.408.1216J, 2007AandA...467.1209H}, and the ``Turtle nebula''  with PSR J0357+3205 \citep{2013ApJ...765...36M}. The first two display bizarre extended features {\em orthogonal} to the pulsar's proper motion directions.  The third PWN represents a long  and luminous tail in the direction opposite to that of the pulsar's motion; however,  close to the pulsar the tail is very  faint (undetectable) with no sign of a bright ``head'' (or compact nebula) near the pulsar (cf.\ Mouse, B0355+54, or J1509--5058 PWNe).  Similar puzzling behavior is seen for  PSR J1101-6101 in the  Lighthouse PWN. To explain these structures, several hypothesizes have been suggested. Sideways structures in PSRs J1101-6103 and   B2224+65 could  be pulsar jets (e.g., \citealt{2010MNRAS.408.1216J, 2014AandA...562A.122P}), although the one-sidedness and  high X-ray efficiencies ($L_X/\dot{E}$) of these structures remain puzzling. The leakage of the wind particles from the apex of the bow shock  pushed too close to the termination shock may be an alternative possibility \citep{2008AandA...490L...3B}. In the latter case, the morphologies  of the sideways features are expected to follow the morphology of the magnetic field in the surrounding  ambient medium (which, in these cases, is the ISM well outside the pulsar's host SNRs).

The bowshock PWNe with tails are  often found around older pulsars moving fast through the rarefied ISM  (perhaps due to the observational bias; see KP08).  Unlike torus-jet PWNe,  the bowshocks   are often  seen in H$_{\alpha}$. However, H$_{\alpha}$ bowshocks and bright X-ray tails are rarely seen together (exceptions are PSR J1741-2054, PSRs J2124-3358,  and binary  B1957+20). The scarcity of such cases suggests that the pre-ionization of the oncoming ISM by the high-energy radiation from the compact PWN and/or pulsar can play an important role. The cases with  both   H$_{\alpha}$ and  radio emission are also very rare but this may be due to the limited number of objects observed  sufficiently  deep in the radio. The sizes of extended pulsar tails can reach 7--8 pc in X-rays (e.g., PWN of PSR J1509--5850) and up to $\sim17$ pc in radio (e.g., Mouse PWN). These two PWNe, well studied both in X-rays and radio, show remarkably contrasting behavior. The Mouse PWN appears to be the brightest closer to the pulsar {\em both in radio and X-rays} while the radio emission from the J1509--5850 PWN is lacking near the pulsar and becomes brightest only a few arc minutes away from the pulsar (see Fig. \ref{tailsheads}). This could be attributed to the fact that the Mouse PWN is seen more of a head-on but the very long radio tail argues against the very small angle between the pulsar velocity and the line-of-sight (see Figure 1 in \citealt{2005AdSpR..35.1129Y}). The morphology of the radio PWN of  PSR J1101--6101 is similar to that of J1509--5850 radio PWN. Furthermore, the radio polarization measurements indicate that in the Mouse tail the magnetic field direction is predominantly  parallel to the tail \citep{2005AdSpR..35.1129Y}, in the J1509--5850  PWN it is mostly perpendicular to the tail \citep{2010ApJ...712..596N}. The differences may be related to different angles between the pulsar velocity  and spin axes in these two PWNe. Alternatively a different ambient density and  entrainment rate could play a role. 

PWNe with $H_{\alpha}$ bow shocks  are particularly interesting  because the  $H_{\alpha}$ emission  allows one  to map the structure of the forward shock not only in coordinate space but also in velocity space  through the measurements of the Doppler shifts in hydrogen lines across the forward shock \citep{2010ApJ...724..908R,2014ApJ...784..154B}. For instance, \cite{2010ApJ...724..908R} performed spectroscopic observations for J1741--2054 and measured  the radial velocities up to $\simeq50$ km s$^{-1}$ consistent with the bowshock model implying pulsar speed of  $\sim150$-200 km s$^{-1}$ and inclination angle\footnote{The angle between the line of sight and the pulsar velocity vector.} of about $75^{\circ}$ (see also \citealt{2015arXiv150103225A}).

\begin{figure}
\includegraphics[width=0.49\textwidth]{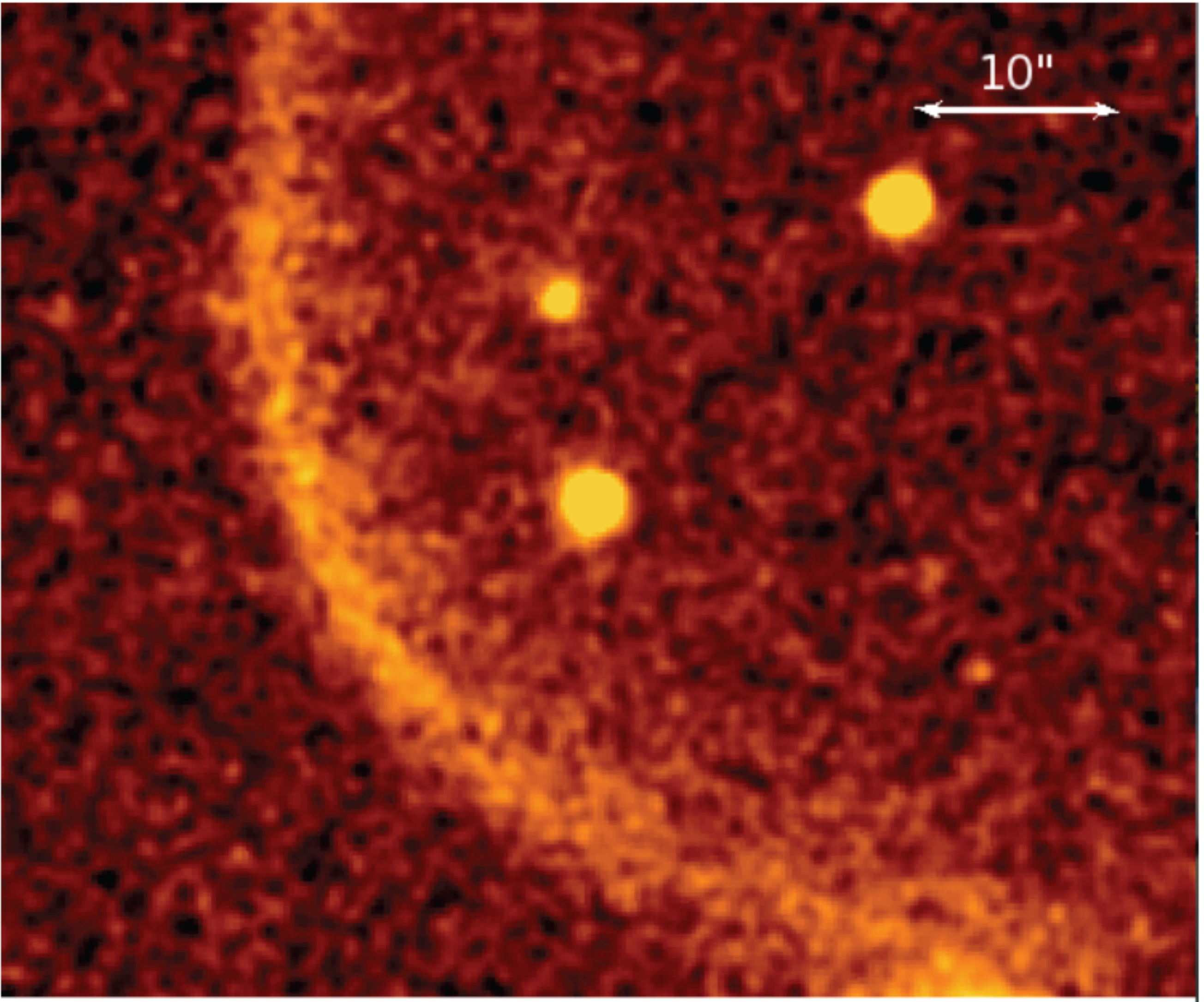}
\includegraphics[width=0.49\textwidth]{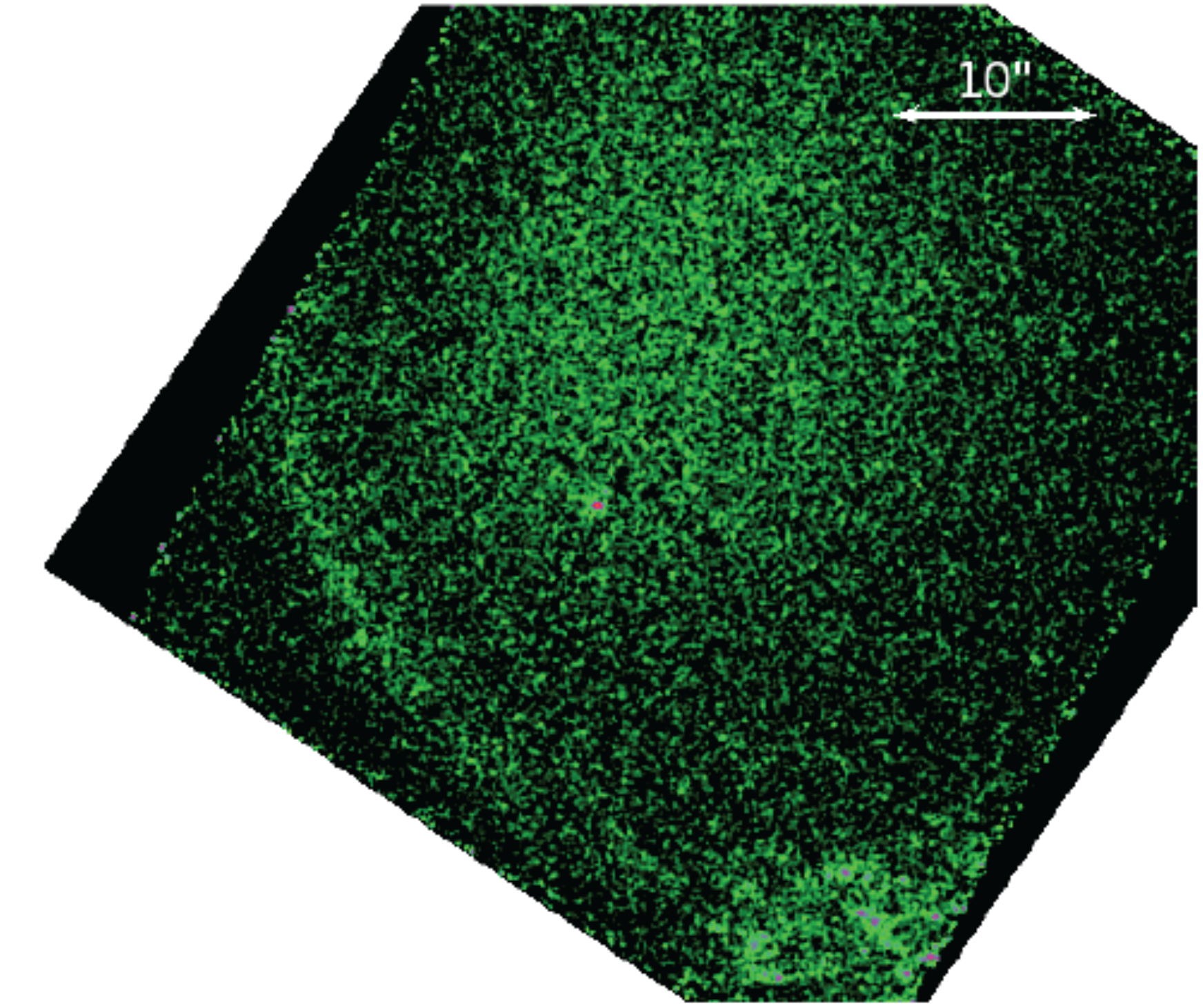}
\caption{ $H_{\alpha}$ (left)  and far-UV (right) images of the bow shock around PSR J0437--4715. Faint, amorphous diffuse emission seen around the pulsar in the right panel is  the instrumental artifact (thermal glow of {\sl HST}/SBC detector) while the structure seen at the bottom is a background galaxy. Both right and left panel images show the same area of sky.  (Rangelov et al.\ in prep.) }
\label{J0437}
\vspace{-0.4cm}
\end{figure}
Measurements of  spectral line ($H_{\alpha}$ or/and $H_{\beta}$) fluxes can  provide accurate diagnostics of the ambient medium density if other parameters (pulsar velocity $V_{\rm psr}$, distance to the pulsar $d$, and the stand-off distance $r_{cd}$) are well constrained. A nearby millisecond  PSR J0437--4715 with $\dot{E}=5.5\times10^{33}$ erg s$^{-1}$ represents an example of such kind. The pulsar, located at the (parallax) distance of $d=156$ pc and  moving with $V_{\rm psr}=134$ km s$^{-1}$ (for the  inclination $i=53^{\circ}$ inferred from the bowshock shape), shows a prominent H$_{\alpha}$ bow shock (Fig.~\ref{J0437}) with a stand-off distance  $r_s=2.2\times 10^{16}$ cm. This implies ISM density $n_{H}=\dot{E}/4\pi c r_{s}^2 m_H V_{\rm psr}^2 =0.1$ cm$^{-3}$. For slow ($V_{\rm psr}<10^{3}$ km s$^{-1}$) bowshocks the H$_{\alpha}$ yield (i.e., number of H$_\alpha$ photons per incoming neutral) is $\epsilon_{H_{\alpha}}=0.6(v/100~{\rm km s}^{-1})^{-1/2}$  according to \cite{2007ApJ...654..923H} and the H$\alpha$ flux $f_{H_{\alpha}}\approx0.0074(v/134~{\rm km s}^{-1})^{1/2}(r_s/2.2\times 10^{16}~{\rm cm})^2(d/156~{\rm pc})^{-2}(n_{H}/0.1~{\rm cm}^{-3})\xi_{HI}$ photons cm$^{-2}$ s$^{-1}$ (see e.g., \citealt{2002ApJ...575..407C,  2014ApJ...784..154B}). The measured  H$_{\alpha}$ flux  from the J0437--4715's bow shock apex, $f_{H_{\alpha}}=6.7\times10^{-3}$ photons cm$^{-2}$ s$^{-1}$ \citep{2014ApJ...784..154B}, implies that the neutral $H$ fraction $\xi_{HI}\approx 0.9$. This is somewhat surprising because the classical Str\"omgren radius for the ionizing radiation produced by the pulsar appears to be much larger than the distance to the H$_{\alpha}$ bowshock apex indicating that the classical formula is probably inapplicable when ionizing radiation is X-rays and the ionizing source is moving fast \cite{2001A&A...380..221V, 2015arXiv150607282L}.   PSR J0437--4715 bow shock is the only one from which the far-UV (FUV; see Fig.~\ref{J0437}) emission has been detected (other pulsars with  H$_{\alpha}$ bowshocks may simply be too far, so FUV photons are easily  absorbed). The measurement of the FUV spectral slope, $\alpha\sim 1.5$ (for $F_{\nu}\propto\nu^{-\alpha}$), suggests  either a non-thermal emission mechanism (synchrotron?) with radiation coming from the vicinity of the forward shock (FUV and H-alpha bow-shocks coincide within the measurements errors, $\simeq 0.5''$) or line emission (with multiple unresolved lines resulting in an effectively flat PL spectrum).
  No firm detection of X-ray emission has been reported for the pulsar wind of J0437--4715  yet, although an analysis of  archival {\sl Chandra} ACIS data indicates that a compact, $r\simeq2''-3''$, asymmetric (most of the emission is ahead of the pulsar) PWN with the luminosity of  $3.8\times10^{28}$ erg s$^{-1}$ may be present (Rangelov et al.\  in prep.).

So far B1951+32, J1509--5850, and J2124--3358 are the  only three solitary pulsars where both X-rays and H$_{\alpha}$  have been detected. However, in J1509--5850, H$_\alpha$ emission is believed to come from the photo-ionized medium rather than from the bow shock (see  the discussion in  \citealt{2014ApJ...784..154B}) and for B1951+32 the surrounding environment is complex due to possible contribution to H$_\alpha$ from the filaments of the host SNR CTB 80. Therefore, it is important to perform deep H-alpha imaging and spectroscopy of nearby pulsars with X-ray bow shocks to study the connection between the pulsar wind and shocked ambient medium regions.

\subsection{PWNe in binaries.}

Sufficiently energetic non-accreting binary pulsars may be able to power detectable PWNe.  In binary systems, in addition to the pulsar parameters, the PWN morphology and appearance would depend on the  properties of the binary companion and  parameters of the binary orbit.  Even if the companion star is lacking any wind itself (e.g.,  it is a cold white dwarf or old low-B neutron star) the pulsar wind will be affected by the ram pressure due to its own orbital motion if the ram pressure caused by the orbital motion is comparable to or larger than the ambient pressure (or ram pressure due to the  motion of the binary as a whole).  Thus, binary motion can  strongly affect the pulsar wind flow (and morphology of the magnetic field) downstream of the termination shock or even the termination shock itself (see e.g., \citealt{2012A&A...544A..59B, 2013A&A...551A..17Z}). If the companion star has a powerful wind  the interaction becomes even more complex with the outcome  critically dependent on the ratio of momentum fluxes of the two winds $\eta=\dot{E}c^{-1}/\dot{M}_w v_w$ (where $\dot{M}_w$ and $v_w$ are the mass loss rate and the massive star wind velocity),  a key parameter in colliding wind binaries (see \cite{2013AandARv..21...64D}). If $\eta\gg1$, the pulsar wind dominates (e.g., for B1957+20 - the original  ``black widow'' system) while in the opposite case the companion star wind will be dynamically dominant. In general, $\eta$ can vary with the orbital position  (1) if  the pulsar wind (or the massive star wind)  is anisotropic, and the spin axis of the pulsar (or the massive star) do not coincide with the orbital angular momentum vector or (2) if the orbit is highly eccentric.

The most famous example of a pulsar binary system where all these effects  play a significant role is LS 2883 with the young energetic pulsar  B1259--63 (B1259 hereafter) in an eccentric 3.4-year orbit around a massive O-star. Although  direct  observations of pulsar wind in B1259 may not be feasible (except perhaps for the VLBI imaging observations; see \citealt{2012AIPC.1505..386M}), there are  indirect ways to learn about the pulsar and stellar wind properties and their interaction. These include multi-wavelength spectrum and flux measurements  as a function of orbital phase \citep{2014MNRAS.439..432C}, pulsar radio signal variability measurements (\citealt{2014MNRAS.437.3255S} and references therein), and,  a high-resolution  X-ray imaging, which recently revealed a dynamic structure  associated with the binary \citep{2014ApJ...784..124K, 2015ApJ...806..192P}.  An X-ray emitting cloud was found to be moving away from the binary with the velocity of $\approx0.07c$ \citep{2015ApJ...806..192P}, which, together with the lack of deceleration, implies either a hadronic cloud with very large mass   $\gtrsim 10^{27} n (d/2.3~{\rm kpc})^{2}$ g moving in the O-star wind with density $n$ or a lighter cloud moving in the rapidly expanding, {\em unshocked} relativistic pulsar wind (implies $\eta\gg1$ in the polar O-star wind). The former scenario implies that the cloud was ejected from the binary during the 2011 periastron passage, when the pulsar interacted with the excretion disk of the massive O-star.  However, the corresponding kinetic energy of the cloud must be very large, $\sim2\times10^{45}$ erg, and  it  must  have been launched via a complex interaction between the pulsar wind and excretion disk of the O-star with the energy source being problematic (pulsar's $\dot{E}$ can only provide $\sim 10^{42}$ erg during the disk passage). The latter scenario is at odds with the common assumption of $\eta\lesssim1$ for such kind of binaries but it does not require extreme values of mass and energy  for the cloud \citep{2015ApJ...806..192P}.

We note that there are other systems where pulsars (albeit these may not be as young and energetic) might orbit massive (often Be or O type) stars, and it is plausible that other TeV gamma-ray binaries   (e.g., LS 5039, LS I +61 303, and HESS J0632+057) also host pulsars (\citealt{2013AandARv..21...64D, 2014AN....335..301K} and references therein). Indeed, \cite{ 2011ApJ...735...58D} reported evidence for amorphous arcminute-scale X-ray emission with a hard spectrum around LS~5039 which was interpreted as synchrotron radiation from ultrarelativistic (pulsar wind?) particles escaping from the system.

Among other types of binaries which can shed light on the properties of pulsar winds at smaller distances from the pulsar through the interaction with the companion are the famous double pulsar  (see e.g., \citealt{2005ASPC..328...95A}), recently reported very eccentric binary with PSR J2032+4127 \citep{2015arXiv150201465L},  and black widow pulsar B1957+20 \citep{2012ApJ...760...92H}.

\begin{figure}
\includegraphics[width=0.322\textwidth]{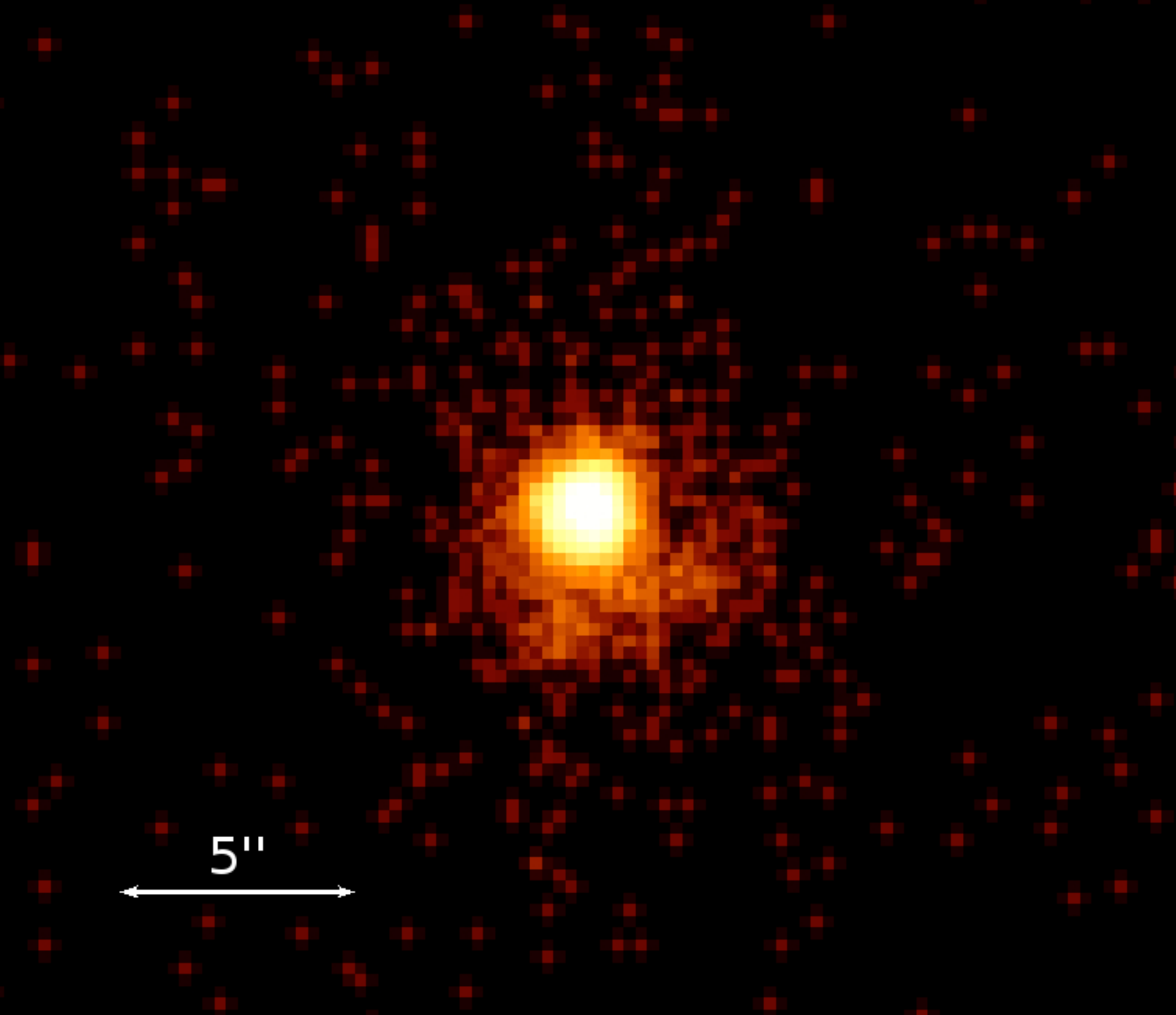}
\includegraphics[width=0.322\textwidth]{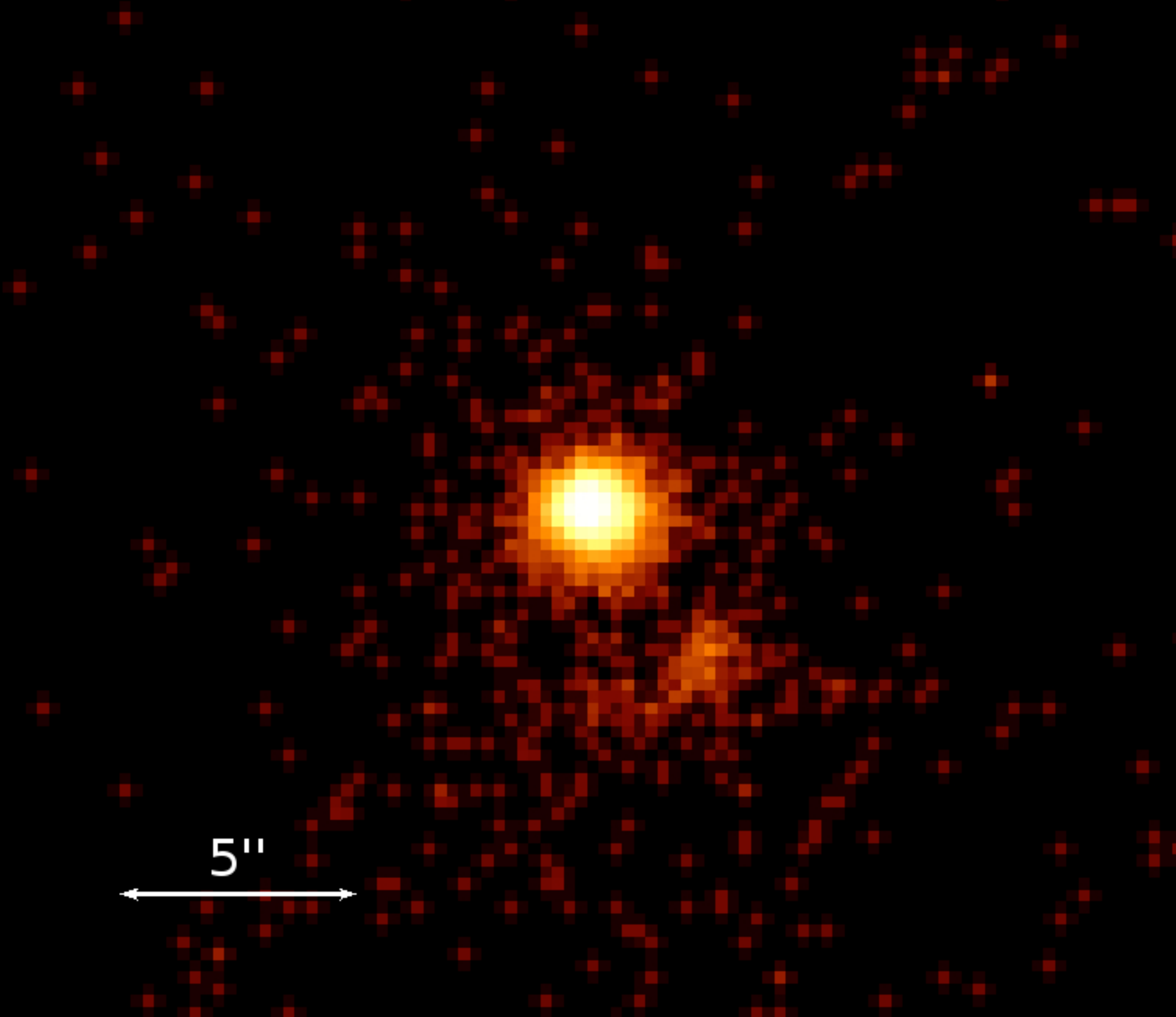}
\includegraphics[width=0.322\textwidth]{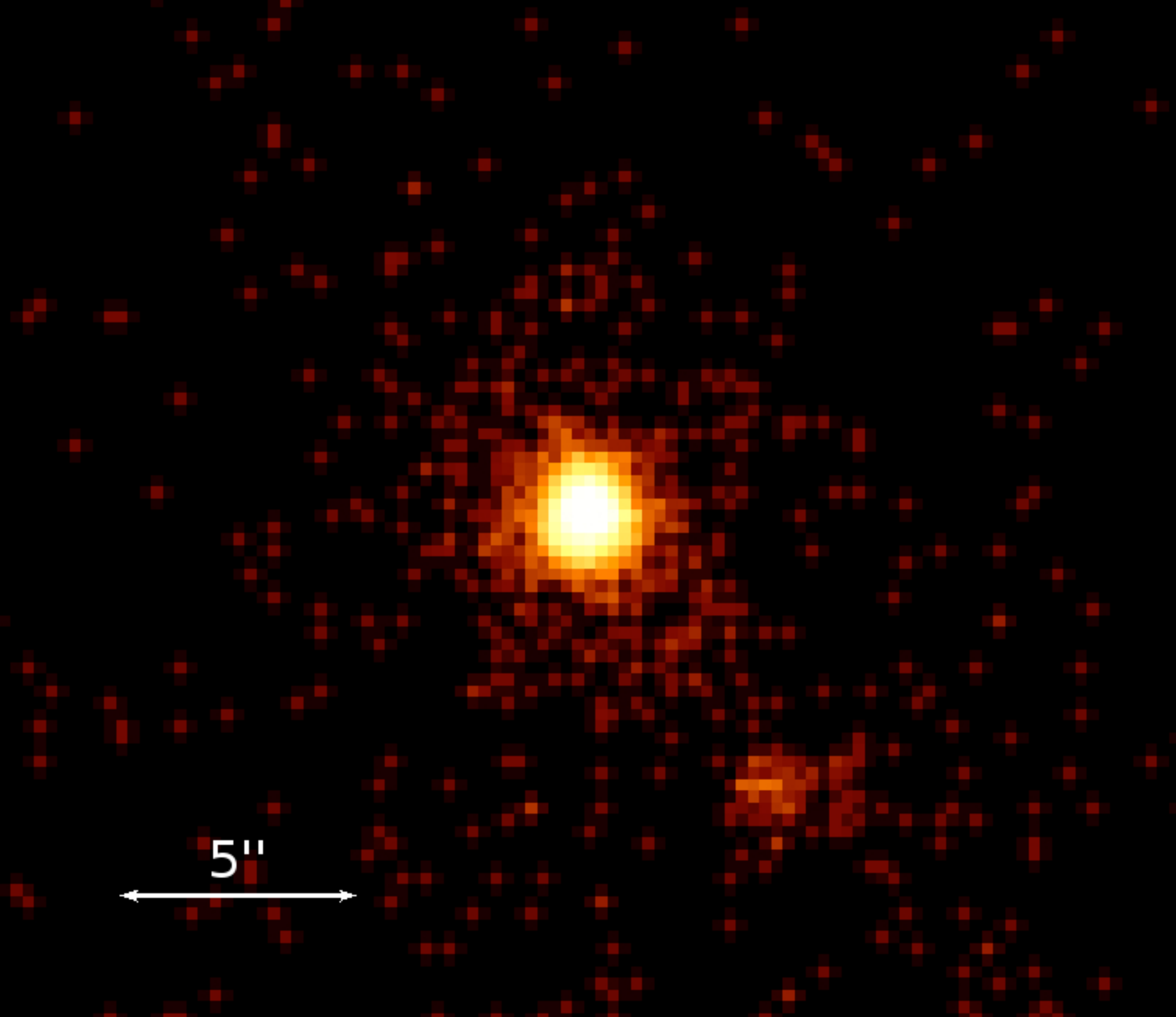}
\caption{  \cxo observations (on MJD 55912, MJD 56431, and MJD 56696 left to right) of LS 2883/B1259-63 reviling the relativistic, $v=(0.074\pm0.006)c$, motion  of  cloud ejected from the binary. See the corresponding movie at \href{http://home.gwu.edu/~kargaltsev/B1259.html}{http://home.gwu.edu/$\sim$kargaltsev/B1259.html}}
\label{b1259}
\end{figure}

\subsection{PWN spectra}

\begin{figure}
\includegraphics[width=0.999\textwidth]{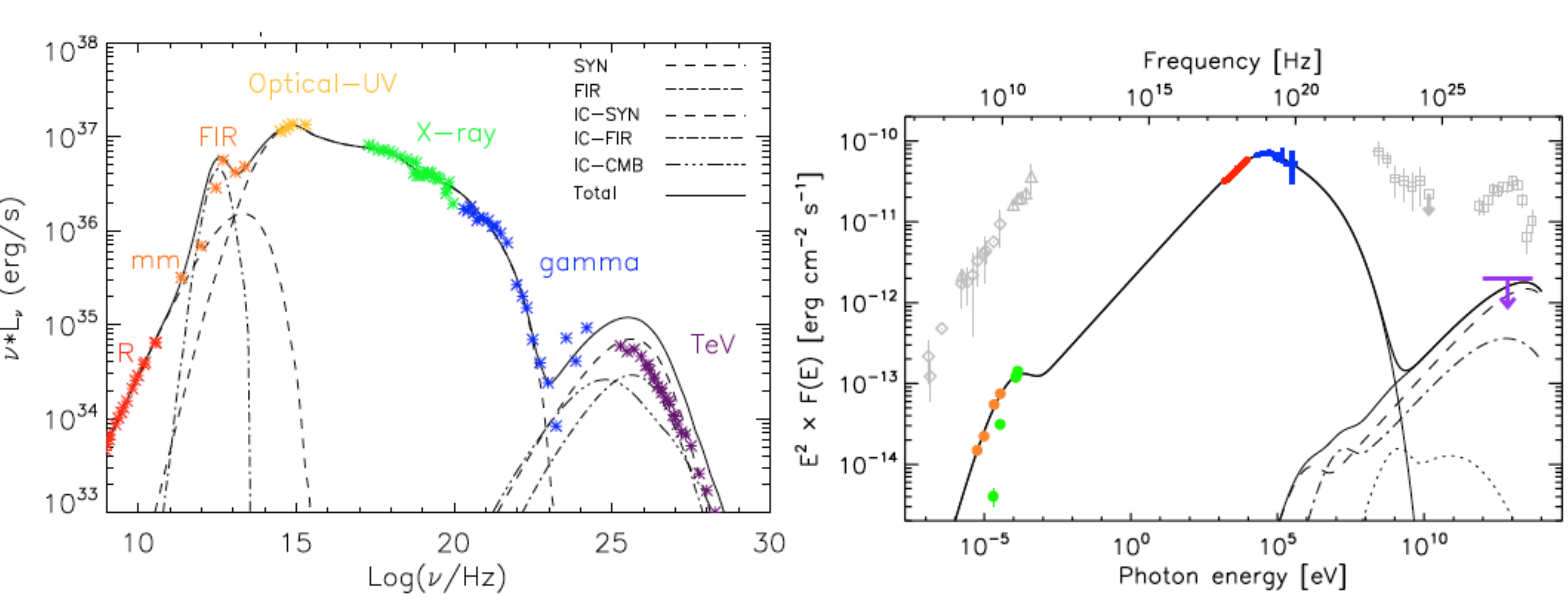}
\caption{ Multiwavelength spectra of the Crab (left; from \citealt{2008A&A...485..337V}) and Vela PWN (right; spectrum from $r=6'$ from pulsar shown in color while the relic Vela X plerion spectrum is in grey; from \citealt{2011ApJ...743L..18M}).}
\label{mwspectra}
\end{figure}

It is commonly accepted that  radio-to-MeV emission in PWNe  is the synchrotron emission. This  naturally explains  the observed high degree of polarization in X-rays, optical, and radio \citep{2013MNRAS.433.2564M,2004ApJ...615..794B,1978ApJ...220L.117W}.  Multiwavelength observations and modeling show that  IC scattering on Cosmic Microwave Background and NIR/IR background starts to dominate  PWN spectra in GeV \citep{2013ApJ...773...77A} and completely dominates synchrotron at TeV energies  (\citealt{1996MNRAS.278..525A}, \citealt{2008A&A...485..337V},  \citealt{2009ApJ...703.2051G}, \citealt{2013MNRAS.436.3112T}, and  \citealt{2014MNRAS.438.1518O}; see the Crab  and Vela PWN  spectra in Fig.\ \ref{mwspectra} for example). The theoretical models of particle acceleration and non-thermal emission in PWNe predict, in the ideal MHD framework, a cutoff around 150 MeV (see \S3.1).
These models well reproduce the observed cutoff in the Crab Nebula spectrum (see left side of Fig. \ref{mwspectra}). However, the slow acceleration rates of ideal MHD models (as, e.g., diffusive shock acceleration) do not allow fast variability in the nebular emission (see  \S3 for Crab flares). The questions about the location and distribution of acceleration sites as well as acceleration mechanism in the pulsar winds and PWNe also remain open (see \S2.4).

If the pulsar wind contains relativistic protons, it is possible that hadronic emission (due to neutral pion decay) contribution can become appreciable in dense environments \citep{2008MNRAS.385.1105B}. Multiwavelength emission from bow shock PWNe (including pulsar tails) produced by pulsars moving in a low-density ISM (outside their host SNRs)  should be purely leptonic. However, surprisingly  few of these objects have been detected in TeV (one of the deepest limits, $0.1\%$ of $\dot{E}$ in 1--10 TeV, is obtained with VERITAS for the tail of B0355+54; Brett McArthur, private communication).   Recent review of the observational X-ray and TeV properties for the population of  91 PWNe can be found in \cite{2013arXiv1305.2552K}.
	
\begin{table*}
\begin{center}
\caption{Photon indices ($\Gamma$) and electron SED slopes ($p=2\Gamma-1$) measured from {\sl Chandra ACIS} data for the innermost regions in 9 bright PWNe \citep{2014cosp...40E1398K}.}
\label{table2} 
\begin{tabular}{rlc}
\hline\noalign{\smallskip}
PWN & $\Gamma$ &  p \\
\noalign{\smallskip}\hline\noalign{\smallskip}
Crab & $1.80\pm0.05$ & 2.6\\
Vela & $1.30\pm0.05$& 1.6 \\
3C58& $1.9\pm0.07$& 2.8\\
G320.4-1.2& $1.4\pm0.1$& 1.8\\
Kes 75& $1.9\pm0.1$& 2.8\\
G21.5-0.9& 1.40$\pm0.06$& 1.8\\
G11.2-0.3& $1.5\pm0.1$& 2.0\\
CTB 80& $1.7\pm0.1$& 2.4\\
G54.1+0.3& $1.50\pm0.05$& 2.0\\
\noalign{\smallskip}\hline
\end{tabular}
\end{center}
\end{table*}

\subsubsection{Spatially resolved X-ray spectra}

Deep {\sl Chandra} ACIS observations of a few bright PWNe allow us to create spectral maps with high-spatial resolution \citep{2013ffep.confE...7K}.  These maps are expected to provide manifestation of synchrotron burn-off (X-ray spectra should become softer farther away from the pulsar) which depends on the strength of magnetic field and bulk flow speed \citep{1984ApJ...283..710K, 2001ApJ...559..275W, 2009ApJ...703..662R} but they may also contain signatures of spatially distributed (in-situ) particle acceleration or rapid particle diffusion.  These spectral maps demonstrate  that the pulsar spectra measured just downstream of the termination shock can differ substantially. From  Table \ref{table2} one can see that although for the Crab PWN the inferred (assuming synchrotron emission model) slope of the electron SED, $p=2\Gamma-1$, is consistent with  the $p=2.1-2.8$ expected from the commonly invoked Fermi acceleration mechanism \citep{2001MNRAS.328..393A, 2009ApJ...698.1523S, 2011ApJ...726...75S},  for the Vela PWN the spectrum is much harder ($p\approx 1.6$) suggesting that a different mechanism might be at work (e.g., magnetic reconnection; \citealt{2014ApJ...783L..21S}).  At least some of the  long pulsar tails (e.g., tail of PSR J1509--5850) tend  to show very little evidence of cooling (in terms of spectral softening in X-rays)  which either suggests ongoing in-situ acceleration along the tail or extremely  fast bulk flow (Klingler et al.\ in prep.).

\subsubsection{X-ray and TeV efficiencies of PWNe}

\begin{figure}
\includegraphics[width=0.99\textwidth]{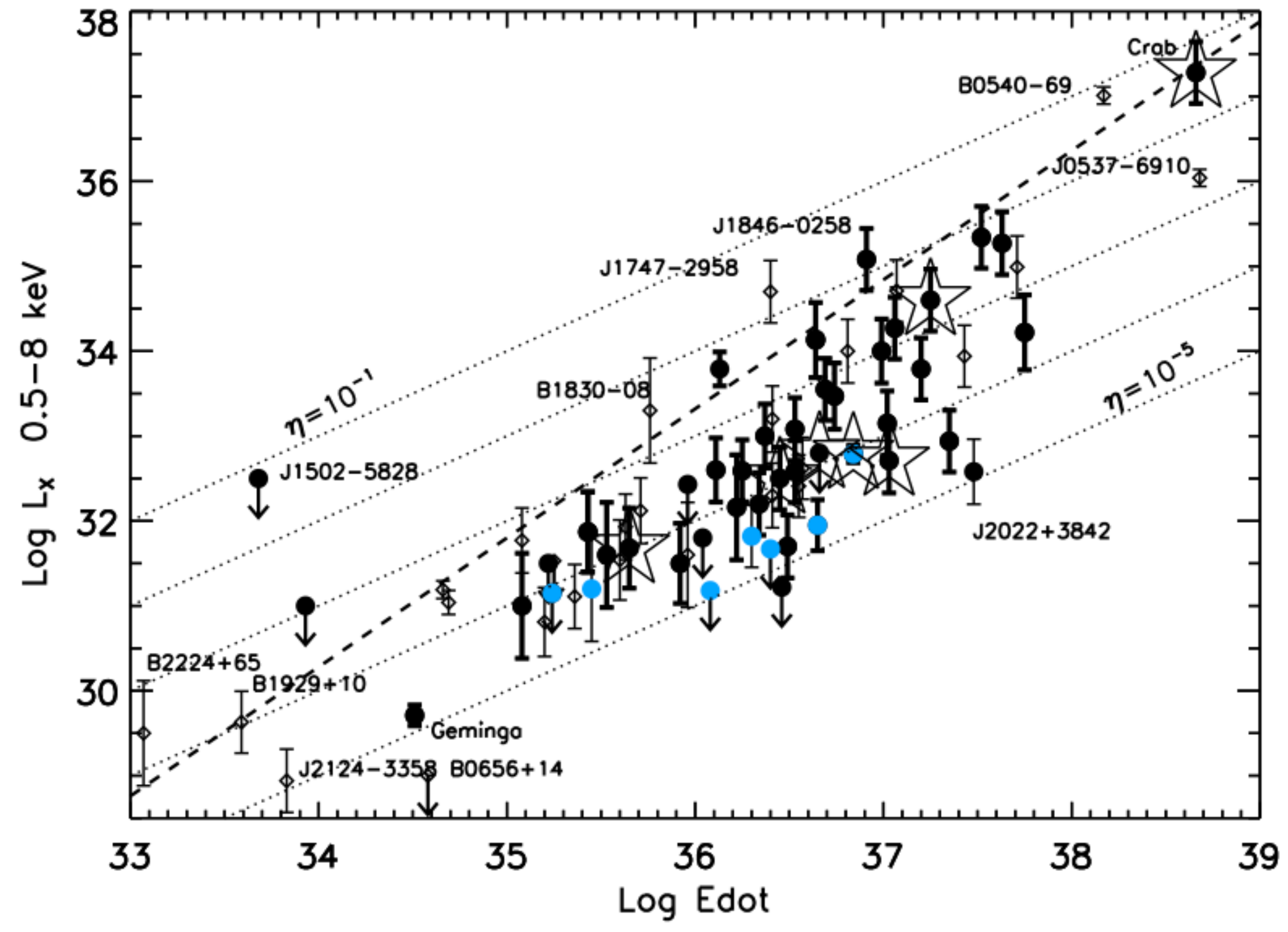}
\includegraphics[width=0.99\textwidth]{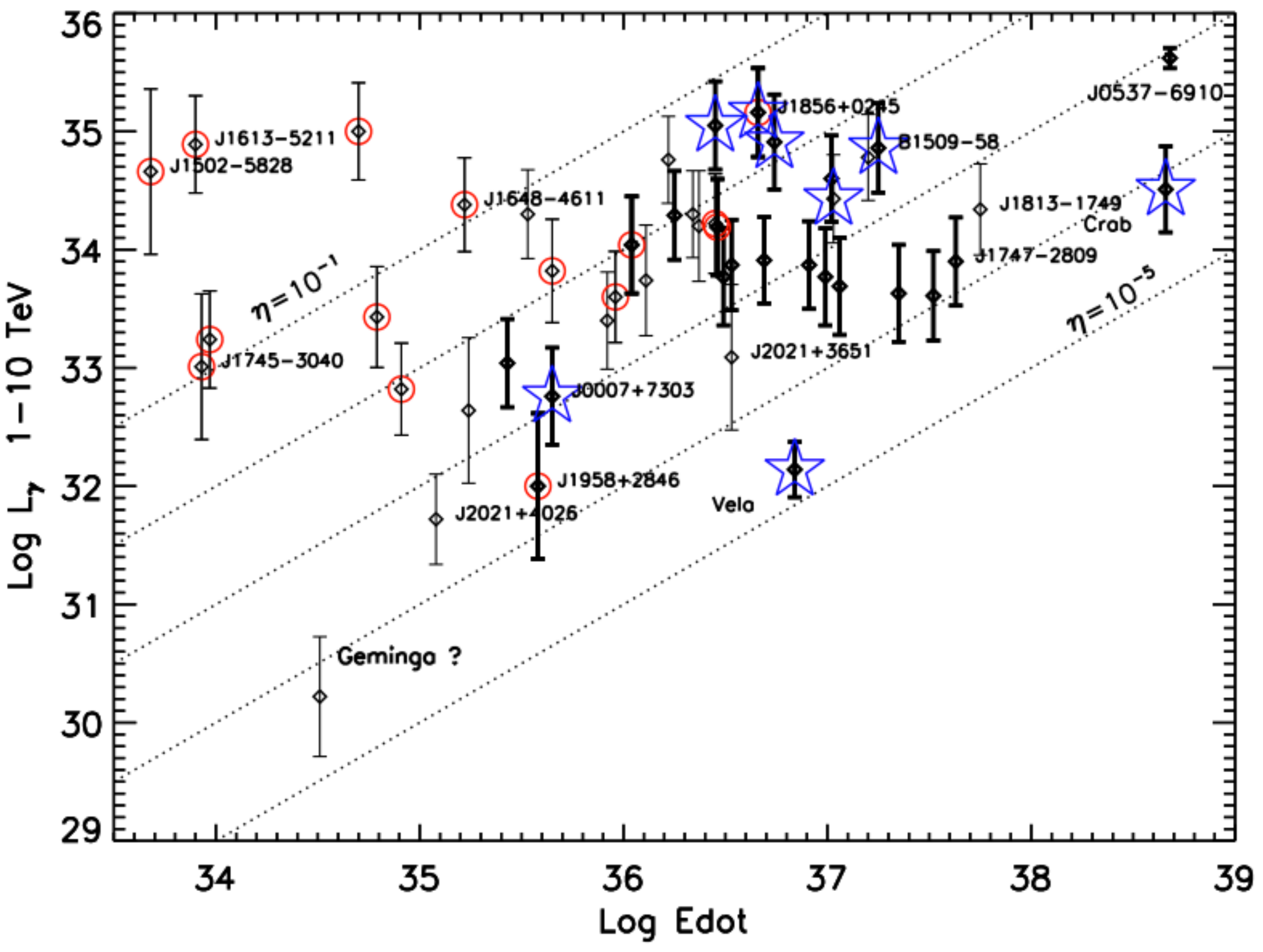}
\caption{{\em Top}: X-ray luminosities of PWNe and PWN candidates vs.\  pulsar's $\dot{E}$. TeV PWNe and TeV PWN candidates  are shown in red. The dotted straight lines correspond to constant X-ray efficiencies; the upper bound, $\log L_X^{\rm cr} = 1.51\log\edot -21.4$, is shown by a dashed line \citep{2012ApJS..201...37K}. The PWNe detected in GeV by {\sl Fermi} are marked by stars. Blue filled circles are the pulsars with confidently measured small ($<10^{\circ}$) magnetic inclination angles from \cite{2014arXiv1410.3310R}. {\em Bottom:} TeV luminosities of PWNe and PWN candidates  vs.\  pulsar's $\dot{E}$.  Thin error bars mark questionable associations. The PWNe undetected in X-rays are marked by circles.  PWNe detected by {\sl Fermi} are marked by stars. The dotted lines  correspond to constant values of the TeV $\gamma$-ray efficiency $\eta_\gamma=L_\gamma/\edot$. The detection of TeV emission from Geminga region is, so far,  based solely on Milagro result \cite{2009ApJ...700L.127A} which has not been confirmed by any other observatories (e.g., VERITAS).   }
\label{pwneff}
\end{figure}

The substantial number of PWNe detected in X-rays and TeV allows  one to investigate the population properties. Here we will only consider X-ray and TeV  radiative  PWN efficiencies ($\eta_{\rm \gamma,X}=L_{\rm \gamma,X}/\dot{E}$)  and refer the readers to \cite{2013arXiv1305.2552K} for the analysis of other TeV and X-ray  properties of PWNe. Figure \ref{pwneff}  (based on the information collected in Tables 1--3 in \citealt{2013arXiv1305.2552K},  with some updates) shows the X-ray and TeV luminosities of PWNe (and PWN candidates  for TeV). Notice a very large spread of X-ray efficiencies and a noticeably smaller spread for the TeV efficiencies.  While there is a clear correlation between $\dot{E}$ and $\eta_{X}$ there is no noticeable correlation in TeV. Finally, the majority of PWNe that are underluminous in X-rays appears to be  around  $\gamma$-ray loud, radio-emitting pulsars with small magnetic inclination angles (based on \citealt{2014arXiv1410.3310R}).

\subsubsection{Search for new PWNe in GeV $\gamma$-rays.}

PWNe are  the most numerous source class that emerged from the H.E.S.S. Galactic Plane Survey \citep{2013arXiv1307.4690C}, and about 100 PWNe or PWNe candidates are known in X-ray. Nevertheless only very few sources are detected in the MeV--GeV range,  because this energy range falls between the tail of the synchrotron emission and the rising part of the IC emission and because of the lack of sensitive MeV instrumentation with good angular resolution. Often,  search for PWN at GeV energies has to be carried out in the presence of $\gamma$-ray-loud pulsar by looking for the off-pulse emission which could come from the PWN \citep[see][]{2013ApJS..208...17A}. In \cite{2013ApJ...773...77A}, a search for GeV emission from 58 TeV PWNe and unidentified sources was performed, with the requirements of (1) good GeV and TeV spectrum connection and (2) extended emission. A total of thirty sources were detected in the GeV range, for energies above 10 GeV; among them, 11 sources are PWN candidates, and three are reliably identified as new PWNe. These new sources are associated with young (age between 1 and 30 kyr) and powerful pulsars with $\dot{E}$ between $10^{36}$ and $10^{39}$ erg s$^{-1}$. It is interesting to study how the multiwavelength properties of the PWNe evolve with the properties of the host pulsar.
It was found in \cite{2013ApJ...773...77A} that there is no correlation between the GeV luminosity and the age and the spin-down luminosity of the pulsars, and the same for the GeV-to-TeV luminosity ratio.
On the other hand, the GeV-to-X-ray luminosity ratio appears to increase with age in agreement with theoretical models(see e.g.,  \citealt{2013MNRAS.436.3112T}).
  Even less is known about PWN emission at the MeV energies due to the poor imaging capabilities of the MeV telescopes. The  dip is expected to occur at these energies in the spectra of  $\sim$10 to $\sim$100 kyr-old PWNe which are most frequently found in X-rays and TeV. Measuring the cut-off energy   of the synchrotron spectrum  in hard X-rays constrains maximum energies of the electrons in PWNe.

\subsubsection{Multiwavelength spectra}

Spatially-resolved spectroscopy is primarily done in X-rays, thanks to the superb resolution of {\sl Chandra}. It remains rather challenging to obtain spatially-resolved spectral measurements at  other frequencies. In the optical-NIR this is primarily due to the  faintness of the PWN emission (except for the Crab PWN) and contamination by various background sources. In the radio, this is challenging because of the interferometric nature of high-resolution observations and difficulties in combing the requirements of imaging both the large- and small-scale structures. At higher energies (MeV, GeV and TeV), the resolution of the existing telescopes is often insufficient to resolve  PWNe, and in those cases when large relic PWNe can be resolved \citep{2006ApJ...636..777A,2012AandA...548A..38A} the limited signal-to-noise ratio typically precludes spatially resolved studies (see, however, \citealt{2006AandA...460..365A, 2011ApJ...742...62V}). Therefore, the MW spectra of most PWNe are, by necessity, spatially-integrated which in many cases introduces bias and systematic uncertainties that are difficult to account for \citep{2011ApJ...742...62V}. Indeed, as one can see from Fig.~\ref{mwspectra}, the MW spectrum of the compact Vela PWN is very different from the
MW spectrum of the relic plerion Vela X (which may be a peculiar one in other respects also, see below). The PWN  luminosities shown in Fig.\  \ref{pwneff} are calculated for very different regions in most PWNe and therefore should be treated as such (i.e.\ cannot be treated as if both X-ray and TeV photons are emitted from the same PWN region). If both TeV and X-ray efficiencies (luminosities) of PWNe are compared to those  derived  from a one-zone PWN model, one should be aware of the limitations of this approach.  Multi-zone models, taking into account both advection and diffusion, appear to be the next logical step in theoretical development and in preparation for Cherenkov Array Telescope  (CTA; see e.g., \citealt{2013APh....43..287D,2015AAS...22533603F}).

As an example (albeit perhaps an unusual one) of a relic PWN one can consider  Vela X.  The nearby  Vela  SNR ($\sim8^{\circ}$ in diameter) has a large region of non-thermal radio emission surrounding  the Vela pulsar (see e.g., \citealt{1998MmSAI..69..919B}).
 One of the brightest radio filaments in Vela X, positioned at the southwest of the pulsar, was detected  in X rays with {\sl ROSAT}, {\sl ASCA}, {\sl Suzaku}, and {\em XMM-Newton}  \citep{1995Natur.375...40M, 1997ApJ...480L..13M,2008cosp...37.2105M,2008ApJ...689L.121L} and, more recently, in GeV $\gamma$-rays with {\sl AGILE} \citep{Pellizzoni2010} and {\sl Fermi}  LAT \citep{Abdo_Vela}  and very high energies (0.5--70 TeV) with HESS \citep{2006A&A...448L..43A} and CANGAROO \citep{2006ApJ...638..397E}. The bright X-ray and VHE emission regions  are positionally coincident (they sometimes referred to as  a ``cocoon''), and have been commonly dubbed   a relic PWN  displaced to the south by the unequal pressure of the reverse shock propagating within the SNR. However,  subsequent  deeper observations with H.E.S.S. \citep{2012AandA...548A..38A} and {\em Fermi} LAT  \citep{2013ApJ...774..110G}  revealed a fainter extended emission whose morphology appears  to correlate with the the   double-lobe, large ($\sim2^{\circ}$ in extent) structure  found at 61 GHz in the WMAP images.   This yields strong support to the scenario where two different populations of electrons are needed to reproduce the radio/GeV halo and the X-ray/TeV cocoon, respectively \citep{2008ApJ...689L.125D}.

\section{Theory of PWNe}

\subsection{General properties of pulsar winds and structure of PWNe}

Theoretical studies of PWNe concentrated on interaction of the relativistic pulsar wind with the surrounding plasma. The morphology of nebulae is described in the scope of magnetohydrodynamic (MHD) models, which proved to be very successful. In order to understand the radiation spectra, one has to develop realistic particle acceleration models; this is a much more difficult task. Here we shortly review the recent development of the PWNe theory; for more comprehensive reviews, see, e.g.  \citet{Kirk_Lyubarsky_Petri09,2012SSRv..173..341A,2014RPPh...77f6901B}.

The general features of PWNe are basically dictated by the physics of the pulsar wind:
\begin{itemize}
\item The pulsar wind is composed predominantly of electron-positron pairs, may be with some admixture of ions. The pair content of PWNe suggests that the pair density in the wind is rather high, more than enough to ensure that the wind could be described as an MHD flow  \citep{1996ApJ...457..253D,Bucciantini_etal11}.
\item The wind is strongly magnetized; most of the energy is transferred, at least initially, as Poynting flux.
\item The wind is highly anisotropic: the Poynting flux is maximal in the equatorial belt and goes to zero at the axis \citep{Michel73,Bogovalov99,Tchechovskoy_etal13}.
\item An obliquely rotating magnetosphere produces variable electromagnetic fields that propagate in the wind as MHD waves; specifically in the equatorial belt, where the magnetic field changes sign every half of period, the so called striped wind is formed.
\end{itemize}

The PWN is in fact a bubble filled predominantly by relativistic particles and magnetic fields; it is inflated by the pulsar wind that continuously pumps into the surrounding medium the energy in the form of relativistic particles and magnetic fields. Within the nebula, the fields and the particles are roughly in equipartition therefore the main question is how the Poynting flux in the pulsar wind is converted into the energy of particles (the so called $\sigma$-problem).

Even though the details of the energy transformation process remain obscure, the general picture is robust: most of the energy is transferred in the wind by alternating electro-magnetic field; and the energy is transferred to the plasma when the alternating field dissipates. This conclusion follows from the strong anisotropy of the MHD wind, in which energy is predominantly transferred in the equatorial belt where the striped wind is formed. Of course the fraction of the energy transferred by alternating fields depends on the angle between the magnetic and rotational axes of the pulsar: an aligned rotator does not produce alternating fields at all whereas the energy from a perpendicular rotator is totally transferred by alternating fields. Due to the strong anisotropy of the pulsar wind, most of the energy is transferred by alternating fields even in a moderately oblique case. \citet{2013MNRAS.428.2459K} calculated the fraction of the energy flux due to alternating fields making use of the split monopole model of the pulsar wind \citep{Michel73,Bogovalov99}, in which the Poynting flux is distributed as $\sin^2\theta$, where $\theta$ is the polar angle. He found that even if the angle between the rotational and magnetic axes is $45^o$, as much as 72\% of the total energy is transferred by alternating fields. In real pulsar winds, this fraction is even larger because the Poynting flux in the wind from a rotating dipole magnetosphere is more concentrated to the equatorial belt than in the monopole wind; according to numerical simulations by \citet{Tchechovskoy_etal13}, the angular distribution of the Poynting flux is close to $\sin^4\theta$ in this case.

Even though the dissipation mechanisms for variable fields in pulsar winds are still debated, one has to stress that the particle Larmor radii within the nebula significantly exceed the wavelength of the waves in the pulsar wind. Therefore these waves could not penetrate into the nebula in any case; even if they survive within the wind, they dissipate at the termination shock front \citep{2003MNRAS.345..153L,2007A&A...473..683P,2011ApJ...741...39S}. Therefore the post-shock flow must be the same as it would be if the dissipation has already been fully completed in the wind. The structure of the nebula is determined by the distribution of the total energy and mean magnetic field in the wind. In the equatorial belt, where most of the energy is transferred, the mean field is weak therefore relatively weakly magnetized plasma is injected into the nebula in this region. At higher latitudes, the magnetic field does not change sign and the variable magnetic fields could propagate in the form of fast magnetosonic waves. These waves efficiently decay as a result of non-linear steepening \citep{Lyubarsky03fms} however, they could not transfer a large fraction of the Poynting flux so that the flow magnetization in this region remains large even after they decay.

Taking into account the above properties of the pulsar wind, the observed morphology of PWNe, and first of all the characteristic disk-jet structure \citep{2000ApJ...536L..81W, 2006ARAandA..44...17G}, is naturally explained \citep{Lyubarsky02}. Namely, the disk is formed by the relatively weakly magnetized equatorial flow, which transfers most of the energy. In the Crab, such a disk is clearly seen in the {\it Chandra} map because the X-ray emitting electrons rapidly loose their energy thus making X-rays a tracer for the freshly injected plasma. In other PWNe, like Vela, the disk may even not be seen at all because the high energy electrons fill a much larger volume. The high latitude flow remains magnetized therefore it is compressed by the magnetic hoop stress to form a jet-like feature at the axis. An important point is that in the highly relativistic, super-sonic (more exactly, super-fast-magnetosonic) wind, the magnetic collimation is inefficient; the "jet" is formed beyond the termination shock where the flow is decelerated. Inasmuch as the pulsar wind is anisotropic, the termination shock is highly non-spherical: it lies much closer to the pulsar in the polar regions than in the equatorial belt; therefore the "jet" appears to originate from the pulsar (see also \citet{Bogoval_Khang02}).

This general conjecture has been confirmed by axisymmetric MHD simulations \citep{2004MNRAS.349..779K, DelZanna04, 2006A&A...453..621D, 2008A&A...485..337V, Bucciantini_etal11}. These simulations were able to explain also nontrivial features of the fine structure, such as a mysterious knot, which is located within 1" from the Crab pulsar \citep{1995ApJ...448..240H}. Namely, the knot is a Doppler-beamed emission from the patch of the highly oblique termination shock where the post-shock flow is still
relativistic and directed towards the observer \citep{2004MNRAS.349..779K}. Simulations with a better resolution and therefore with a lower numerical viscosity \citep{2009MNRAS.400.1241C} revealed bright fine filaments moving away from the termination shock with a good fraction of the speed of light. These are highly reminiscent of the so-called wisps of Crab nebula \citep{1995ApJ...448..240H}.

In spite of all these successes, important basic problems have remained unresolved. Namely, axisymmetric simulations reproduce the observed structure of the nebula only if the wind magnetization at high latitudes was chosen to be relatively small, significantly smaller than one could expect from theoretical considerations. This discrepancy was resolved only recently when fully 3D simulations have been performed \citep{2014MNRAS.438..278P}. Let us describe these recent developments in more details.

\subsection{The $\sigma$-problem and 3D simulations of PWNe}\label{sect_sigmapb}

The problem of the magnetic to the plasma energy transformation in pulsar winds is generally referred to as the $\sigma$-problem because the flow magnetization, defined as the ratio of the Poynting to the plasma energy fluxes, is typically denoted by $\sigma$. The pulsar wind starts as a highly Poynting dominated ($\sigma\sim 10^4 - 10^6$); on the other hand, there is a pervasive belief that one can account for the morphology of PWNe, including the remarkable jet-torus structure, only if just upstream of the termination shock, $\sigma$ does not exceed 0.01. Such a tremendous drop in the flow magnetization looks so mysterious that the problem was sometimes  referred to as the $\sigma$-paradox. However, one has to stress that what we really need to consider is the mean field because alternating fields inevitably decay. As it was mentioned above, they transfer most of the energy in the pulsar wind  therefore the magnetization due to the mean field is not large, which makes the $\sigma$-problem not so severe. Let us discuss the issue in a bit more details.

First of all one has to stress that the  strong constraints on the wind magnetization at the termination shock mentioned above were obtained in spherically or axially symmetric models of PWNs \citep{Rees_Gunn94, 1984ApJ...283..694K, emmeringchevalier87, begelmanli92}. The reason for the required low value of $\sigma$ is that in these models,  the magnetic field strength grows with radius within the nebula so that the field  would exceed the equipartition value and pinch the flow too much if the magnetization at the termination shock is not extremely small. The behavior of the magnetic field could be easily understood if one takes into account that the field in the far zone of the wind is practically azimuthal, and in the axisymmetric flow, the field lines remain coaxial circular loops. The radius of the field line increases when the flow expands, the field strength being determined by the conservation of the magnetic flux within the toroidal magnetic tube. At the termination shock, the flow compresses so that the magnetic field increases three times. The flow within the nebula is subsonic therefore the pressure and the density of the plasma do not change significantly. Therefore the volume of the toroidal magnetic tube remains roughly constant. In this case, the cross section of the tube decreases when the tube radius increases, which implies an increase in the magnetic field roughly linearly with the radius.

Taking into account that the size of the nebula is about an order of magnitude larger than the radius of the termination shock, one finds that in the axisymmetric flow, the field strength in the main body of the nebula exceeds that in the wind just upstream of the shock $\sim 3\times 10=30$ times whereas $\sigma$ grows roughly three orders of magnitude. The problem can be alleviated if the kink instability destroys the concentric field structure in the nebula \citep{1998ApJ...493..291B}. Then the magnetic loops could come apart and one expects that in 3D, the mean field strength is not amplified much by expansion of the flow. In this case, $\sigma$ just upstream of the termination shock might not need to be so unreasonably small as was found in axisymmetric models.

This idea can be checked only by 3D simulations of plasma flow within the nebula. As the
first step, \citet{2011ApJ...728...90M} simulated the 3D evolution of a simple cylindrical model of PWNe developed earlier by \citet{begelmanli92}. This model describes a static cylindrical configuration with a relativistically hot plasma such that  the thermal pressure is balanced by the hoop stress of a purely toroidal magnetic field. The simulations clearly demonstrated that the kink instability does develop in the system and destroys the regular concentric structure of the magnetic field thus relaxing the hoop stress and triggering magnetic dissipation. This proves that 3D effects play crucial role in the evolution of PWNe. However, these simulations do not claim to model PWN, simply because the continuous injection of magnetic flux and energy into PWN by their pulsar winds is not accounted for.

The first realistic 3D simulations of  PWNe were performed by \citet{2014MNRAS.438..278P}. They used qualitatively the same setup as in 2D simulations, namely, the nebula is pumped by a strongly anisotropic pulsar wind with the magnetization determined only by the mean magnetic field as if the alternating component of the field has completely dissipated. The difference was in the magnetization at high latitudes, where the magnetic field does not change sign. In 2D simulations, the observed morphology was reproduced only if the  high latitude $\sigma$ was chosen to be as small as 0.1 even though according to the pulsar wind theory, it should remain significant, not less than a few.  \citet{2014MNRAS.438..278P} took $\sigma=1\div 3$ at high latitudes, as it should be.

\begin{figure}
\centering
\includegraphics[width=0.75\textwidth]{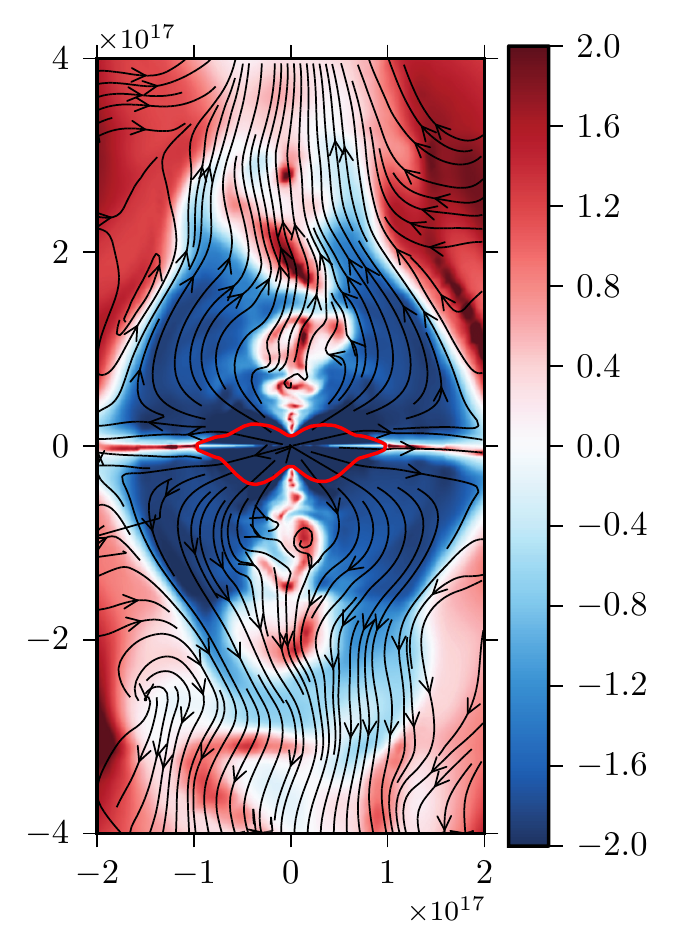}
\caption{Formation of the polar outflow in the 3D MHD simulations by \citet{2014MNRAS.438..278P}. The colour images show the distribution of $\lg\beta$ in the yz plane, where $\beta$ is the ratio of the thermal to the magnetic pressure. The black lines show the momentary streamlines and the
red line the termination shock.}
\label{fig_porth1}
\end{figure}

\begin{figure*}
\centering
\includegraphics[width=0.75\textwidth]{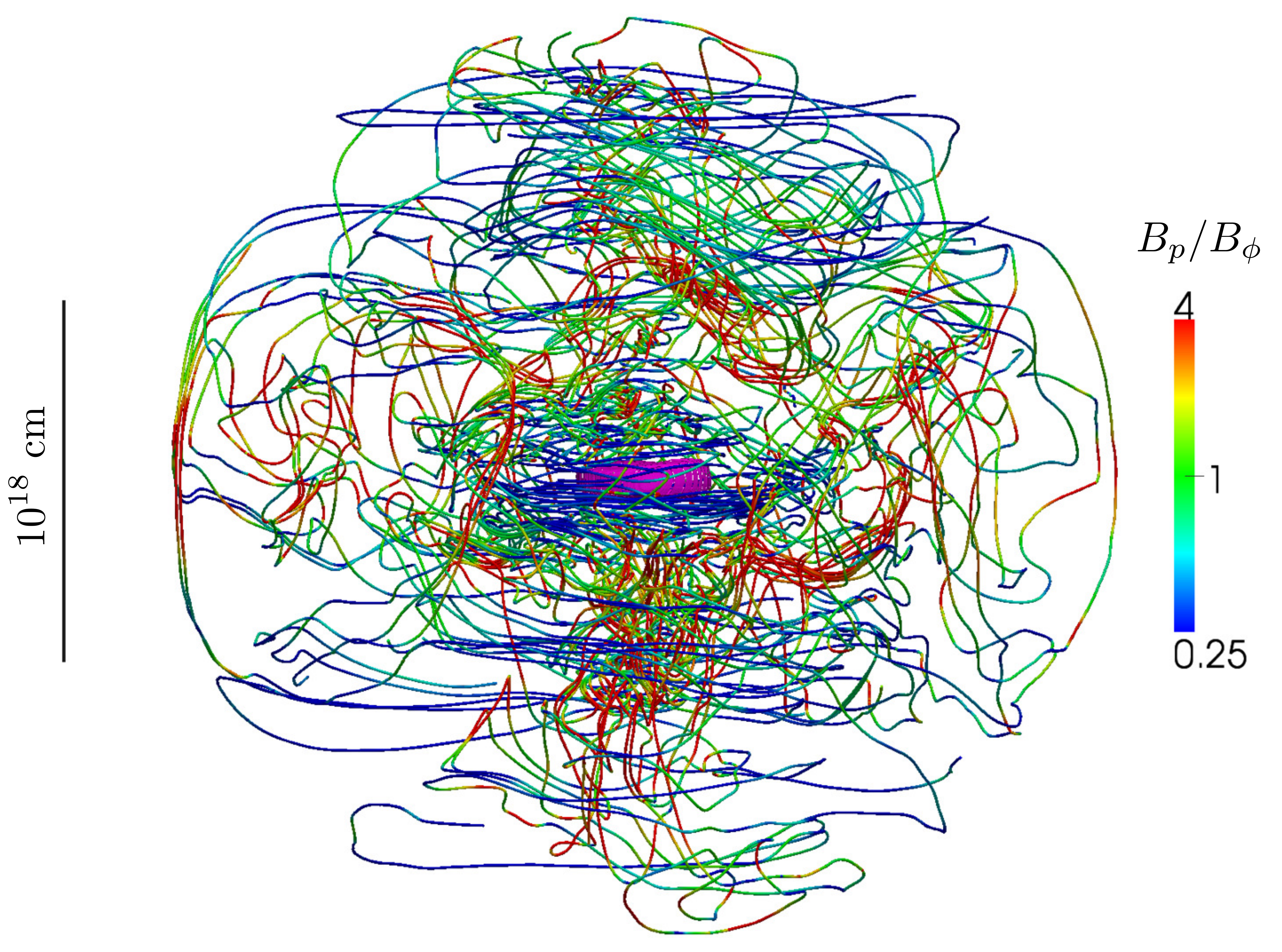}
\caption{Field lines in the 3D simulations by \citet{2014MNRAS.438..278P}. The
lines are coloured according to their orientation, sections with dominating
azimuthal component being blue and those with dominating poloidal component being
red. The surface of the termination
shock is also shown, using the magenta contour.}
\label{fig_porth2}
\end{figure*}

\begin{figure*}
\centering
\includegraphics[width=0.75\textwidth]{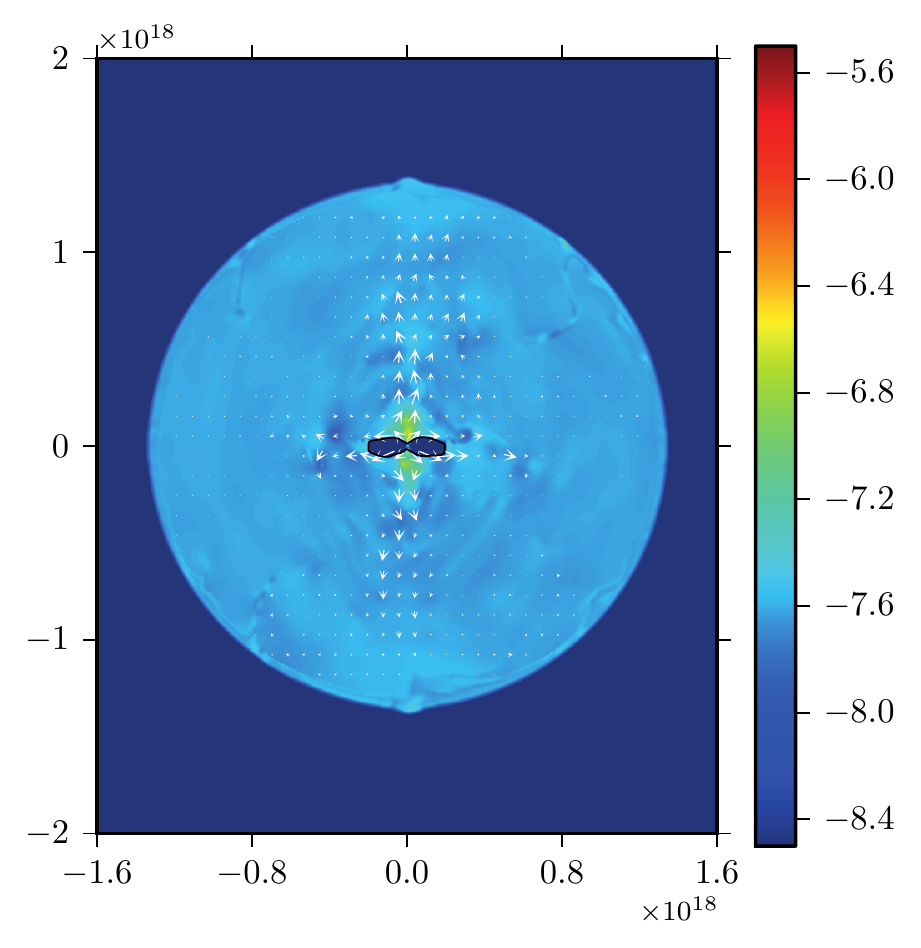}
\includegraphics[width=0.75\textwidth]{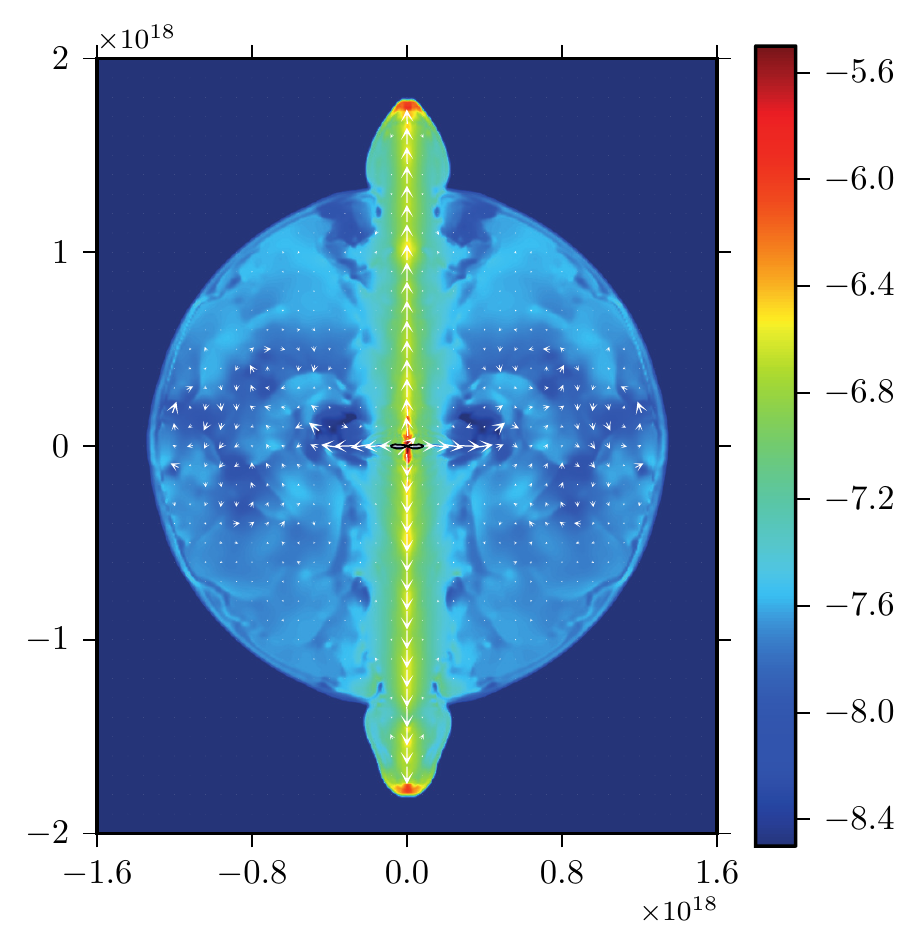}
\caption{Dependence of the total pressure distribution, $\lg p_{tot}$, on the imposed symmetry \citep{2014MNRAS.438..278P}. The upper panel shows the pressure distribution in the
$xz$ plane of the 3D simulations and the lower panel in the corresponding 2D run.
One sees that the strong axial compression observed in 2D simulations is an artifact of the imposed
symmetry.}
\label{fig_porth3}
\end{figure*}

According to the results of their 3D simulations, the azimuthal component of the magnetic field is still dominant in the inner part of the nebula,  which is filled mainly with freshly injected plasma. The hoop stress of this field is still capable of producing noticeable axial compression close to the termination shock and driving polar outflows, required to explain the Crab jet, and jets of other PWNe (Fig.~\ref{fig_porth1}). However, these are much more moderate than in 2D models. In the main body of the nebula, the highly organized coaxial configuration of magnetic field is largely destroyed by the kink instability (Fig.~\ref{fig_porth2}) therefore the global evolution of the PWN in 2D and 3D cases differs radically (Fig.~\ref{fig_porth3}). If the high latitude magnetization is large, the 2D models develop extremely strong polar jets, which burst through the supernova shell. In contrast, in the 3D models the $z$-pinch configuration is destroyed by the kink instability so that the polar outflows are less powerful and eventually lose collimation, as observed.


Important observational constraints are imposed by polarization measurements that reveal a high degree of polarization in the central part of the  Crab nebula with the polarization vector parallel to the pulsar axis, which may be considered as direct evidence for the azimuthal field \citep{2008ARAandA..46..127H}. The simulations do show that in spite of the strong disruption of the azimuthal magnetic field, the polarization remains substantial, particularly in the inner part of the nebula. The polarization direction on the scale of the torus clearly indicates an azimuthal field because the photon magnetic vector appears to curve around the torus.

The 3D simulations also show a bright knot, which was discovered in the previous 2D simulations \citep{2004MNRAS.349..779K} and identified with the inner knot observed in the Crab nebula \citep{1995ApJ...448..240H}. This emission comes from the immediate vicinity of the termination shock, it is highly Doppler beamed and originates in the high-speed part of the post-shock flow. A correlation was found between the knot position and the flux, such that brighter states correspond to a smaller offset between the knot and the location of the pulsar, which is in excellent agreement with the recent optical observations of Crab's knot \citep{2013MNRAS.433.2564M}. The simulated polarization degree and polarization angle in the knot also agree with observations.

The termination shock is found to be unsteady due to an intricate feedback mechanism between
the shock and the nebula flow. The inhomogeneities, formed in the post-shock flow as a result of this variability, appear as wisps emitted from the shock location, in a qualitative agreement with the observations of the Crab nebula.

In the main body of the nebula, the kink instability not only destroys the regular magnetic field structure; the excited turbulence yields efficient magnetic dissipation. In simulations, this occurs at the grid scale via numerical resistivity. To become efficient, it requires creation of ever smaller scale structures in the magnetic field distribution. However, it is important that the processes which drive the development of such structures occurs on scales above the grid scale so that the dissipation rate is determined sufficiently accurately. The simulations with doubled resolution show the same dissipation rate, which suggests that the high degree of dissipation observed in these simulations is not far from being realistic. This also agrees with the observations of synchrotron and inverse-Compton emission of the Crab nebula, which show that the magnetic field is energetically sub-dominant to the population of relativistic electrons by a factor of $\sim 30$ \citep{Hillas98}.

The above results suggest that the magnetic dissipation inside PWNe is a key factor of their dynamics. Combined with the magnetic dissipation in the striped zone of the pulsar wind, it allows to reconcile the observations of the Crab nebula with the expected high magnetization of such winds, thus finally resolving the long-lasting $\sigma$-problem.

\subsection{The structure of the highly magnetized region in PWNe}

Even after decay of the alternating fields, the pulsar wind remains highly magnetized at high latitudes. The energy and momentum flux in this domain of the wind is relatively small therefore the termination shock approaches significantly closer to the pulsar near the axis than at the equator. This region is not quite well resolved in simulations; they just show rapid disruption of the flow by the kink instability. On the other hand, the highly magnetized region is of special interest because the recently discovered strong, short gamma-ray flares from the Crab nebula \citep{2011Sci...331..739A, 2011Sci...331..736T} are generally attributed to a rapid magnetic energy release via, e.g., reconnection (see Sect.~\ref{sect_flares}), which assumes a magnetically dominated region within the nebula. \citet{2012MNRAS.427.1497L} developed a simple model clarifying the structure of the high latitude flow.

In a highly magnetized flow, only weak shocks could arise therefore the termination shock at high latitudes is weak; and the postshock flow in this region remains radial and relativistic. \citet{2012MNRAS.427.1497L} found a simple analytical solution for such a relativistic postshock flow. According to this solution, the flow initially expands and decelerates but eventually becomes to converge because the magnetic hoop stress is not counterbalanced either by the poloidal field or by the plasma pressure. In the converging flow, magnetic energy is converted into the plasma energy therefore the plasma accelerates and heats. If the flow remained axisymmetric, it would eventually be focused at the axis, the magnetic energy being transferred to the plasma. The focus occurs on the axis of the system at the distance from the pulsar $\sim\theta_0^2a$, where $\theta_0$ is the opening angle of the highly magnetized part of the wind, $a$ the equatorial radius of the termination shock. This point may be identified with the base of the observed jet.

An important point is that in a converging flow, even small perturbations eventually destroy the regular structure. The reason is that if converging loops are initially shifted one with respect to another by a displacement much less than their radius, the distortion becomes strong when the radius of the loops approaches the initial displacement. One has to conclude that when the axisymmetric flow is focused into a point at the axis, the magnetic loops inevitably come apart close enough to the converging point giving rise to a specific turbulence of shrinking magnetic loops. Hence one can expect that the energy of the highly magnetized part of the pulsar wind is released in a small region close to the converging point; this gives rise to the observed jet.  Relativistic turbulent motions in highly magnetized plasma imply $E\approx B$ so that in the energy release region, particles could be efficiently accelerated either via the second order Fermi mechanism or via the magnetic reconnection. Therefore the synchrotron gamma-ray emission in the hundreds MeV band, both persistent and flaring, could come from a small region at the base of the jet.

\subsection{The unsolved problem: origin of PWN spectra}

One sees that the overall morphology of PWNe is now well understood in the scope
of MHD models. However, our ignorance of the physical processes giving rise to particle acceleration forces us to treat the injection particle spectra in PWNe as free parameters \citep{2006A&A...453..621D, 2008A&A...485..337V, 2014MNRAS.438.1518O}, and this freedom in interpreting the data limits the level of scrutiny to which MHD models can be subjected. The radiation spectrum carries information about the particle acceleration processes. Most of the observed radiation (from the radio up to a few hundred MeV) is synchrotron emission, with only the peak in the very high energy gamma-ray band being attributed to the inverse Compton scattering of synchrotron photons off high-energy electrons and positrons. The observed strong polarization in the radio, optical and X-ray bands is a supporting evidence for the synchrotron origin of the nebula emission.

The synchrotron part of the spectrum may be described as a broken power-law. The generic observational feature of PWNe is a flat radio spectrum, ${\cal F}_{\nu}\propto \nu^{-\alpha}$, with $\alpha$ between $0$ and $0.3$, extending in some cases out to the infrared. At high frequencies, the spectrum softens, and in the X-ray band, $\alpha>1$. Such an injection spectrum suggests an unusual acceleration process. The observed radio spectrum implies a power-law energy distribution of injected electrons, $N(E)\propto E^{-\kappa}$, with a shallow slope $1<\kappa<1.6$. Such an energy distribution is remarkable in that most of the particles are found at the low energy end of the distribution, whereas particles at the upper end of the distribution dominate the energy density of the plasma.  Specifically in the Crab Nebula, the observed emission spectrum implies that the particles in the energy range from $E_{\rm min}<100$ MeV to $E_{\rm break}\sim 1$ TeV are injected into the nebula with a spectral slope $\kappa=1.6$, so most of the injected energy ($\sim 5\cdot 10^{38}$ erg/s) is carried by TeV particles, whereas $\sim 100$ times more particles are found at low energies of less than 100 MeV. This means that the acceleration process somehow transfers most of the total energy of the system to a handful of energetic particles, leaving only a small fraction of the energy for the majority of the particles. This is not what one would normally expect from the conventional first-order Fermi acceleration process, in which the particle flow is randomized at the shock and only a fraction of the upstream kinetic energy is deposited in highly accelerated particles.

\citet{2003MNRAS.345..153L} proposed that the flat energy distribution is formed in the course of the particle acceleration by driven reconnection of the alternating magnetic field at the pulsar wind termination shock.  As a model for such a process, \citet{LyubarskyLiverts08}
have performed PIC simulations of driven magnetic reconnection in a pair plasma. Two stripes
of opposite magnetic polarity were compressed by means of an external force, which would imitate the effect of a shock. They found that driven magnetic reconnection can produce flat non-thermal particle spectrum, with $\kappa\approx 1$. Realistic 3D PIC simulations of the shock in a striped wind were performed by \citet{2011ApJ...741...39S}. They found that the spectrum of accelerated particles depends on the parameter $\xi=\lambda/ \sigma r_l$, where $\lambda$ is the stripe wavelength, $r_L=mc^2\Gamma/eB$ the Larmor radius corresponding to the upstream Lorentz factor of the flow, $\Gamma$, and the upstream magnetic field, $B$. It turns out that broad particle spectra with flat slopes ($1<\kappa<2$)  could be formed by the shock-driven reconnection only if the above parameter is not less than a few hundreds. In the opposite case, the spectrum resembles a Maxwellian distribution.

Note that $\sigma\Gamma$ is in fact the Lorentz factor the particles would achieve if the whole spin-down energy were equally distributed between them. Therefore the parameter $\xi$ is in fact equal, to within a numerical factor, to the ratio of the pulsar light cylinder radius to the Larmor radius acquired by the particles when a significant fraction of the Poynting flux is converted to the plasma energy. The latter is generally very large because the magnetic field at the pulsar wind termination shock is weak therefore $\xi$ could hardly exceed unity. The condition for the formation of the non-thermal tail could be achieved only if the pulsar wind is overloaded by pairs; then the energy per particle may be small enough so that the Larmor radius remains small. At present, there is no reason, neither observational not theoretical,  to believe in such an necessary extraordinary large pair production in pulsars. Therefore the problem of flat particle energy spectra in PWNe remains unsolved.

\section{Implications of the Crab flares}\label{sect_flares}

The rapid variability is now a well-established intrinsic property of the Crab Nebula in the GeV gamma-ray band\footnote{The X-ray flux of the Crab Nebula is also variable but to a $\sim 10\%$ level over a year-timescale \citep{2011ApJ...727L..40W}. This will not be discussed here, because this phenomenon does not appear to be directly connected to the flares.} \citep{2011Sci...331..736T, 2011Sci...331..739A}. The flares were not predicted by the models and they generally do not fit in the framework of the classical theory of pulsar wind nebulae and particle acceleration. This unexpected phenomenon is also a challenge for observers, because the Nebula is routinely used as a standard candle for cross-calibrating X-ray and gamma-ray instruments. We explain why the flares are so challenging for the models of the Crab Nebula in Sect.~\ref{sect_puzzle}, and we briefly review some of the current attempts to model the flares in Sect.~\ref{sect_model} (see also the reviews by \citealt{2012SSRv..173..341A, 2014RPPh...77f6901B}).

\subsection{The puzzling features of the flares}\label{sect_puzzle}

\begin{figure}[]
\centering
\includegraphics[width=12cm]{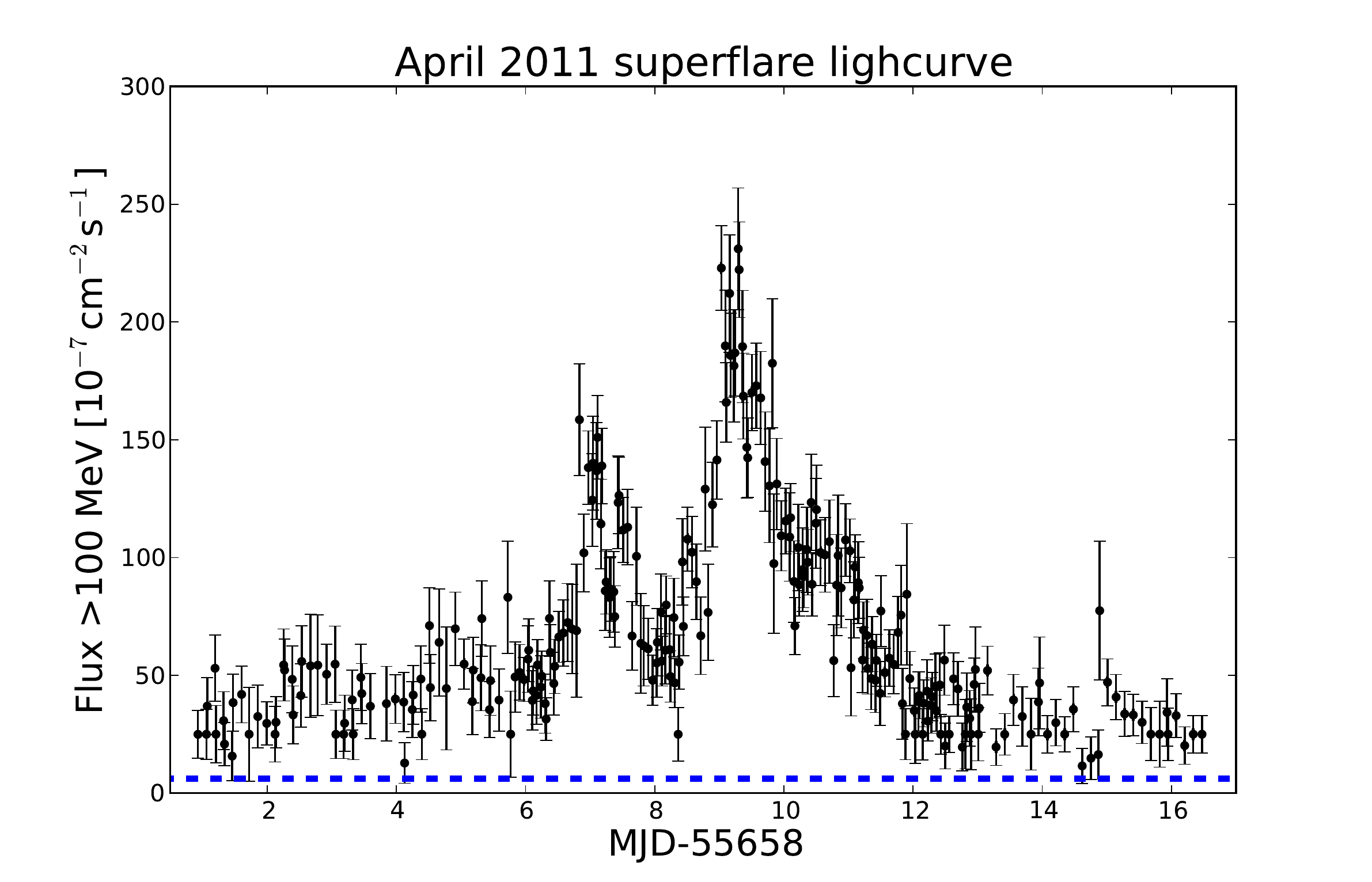}
\caption{Gamma-ray lightcurve above $100$~MeV of the April 2011 super-flare, measured by the {\rm Fermi}-LAT \citep{2012ApJ...749...26B}. The horizontal dashed blue line shows the ``quiescent'' synchrotron flux $>100$~MeV.}
\label{fig_lc}
\end{figure}

With more than 6 years of data, we know that the mysterious engine at the origin of the gamma-ray flares turns on about once or twice a year for about a week\footnote{See \url{http://fermi.gsfc.nasa.gov/ssc/data/access/lat/msl_lc/source/Crab_Pulsar}}. Outside of these spectacular events, identified as the ``flares'', the $>100~$MeV lightcurve remains apparently restless with continuous small variations of the flux \citep{2012ApJ...749...26B, 2013ApJ...765...52S}, as if the engine never really switches off. The duration of the flares indicates that the emitting region must be surprisingly small compared with the size of the Nebula. For a typically week-long episode, the length-scale of the accelerator is of order $ct_{\rm flare}\sim 10^{16}~$cm, i.e., much smaller than the size of the termination shock radius, which is of order $0.1~$pc. The brightest events present intra-flare variability timescales as short as about $8~$hrs (see Fig.~\ref{fig_lc}, and \citealt{2011A&A...527L...4B, 2012ApJ...749...26B, 2013ApJ...775L..37M}), which put even more severe constraints on the size of the particle acceleration site. Consequently, a tiny fraction of the Crab Nebula is radiating $\sim 10$ times more flux than the entire quiescent Nebula in the GeV band. During the April 2011 flare, the gamma-ray flux peaked at about $1\%$ of the spin-down power of the pulsar ($L_{\rm sd}=5\times 10^{38}$~erg/s) \citep{2012ApJ...749...26B}. This is putting strong constraints on the energetic budget required to power the flares.

The gamma-ray flare spectrum appears at the high-energy end of the quiescent synchrotron spectrum, and extends up to about 1~GeV. The flaring component is usually attributed to synchrotron radiation emitted by $10^{15}~$eV (or PeV) electron-positron pairs in a $\sim$ milli-Gauss magnetic field. Other radiative processes such as inverse Compton scattering or Bremsstrahlung are far too inefficient to cool down the particles over the duration of the flare\footnote{The flux of the inverse-Compton component above $>100~$GeV remains constant during the flares \citep{2014A&A...562L...4H, 2014ApJ...781L..11A}.}. It was already known that the Crab Nebula accelerates particles up to PeV energies (e.g., \citealt{1992ApJ...396..161D}). What is new, however, is the evidence that such particles are accelerated over such a short timescale (the rise and decay timescales of the flares range between 6 hours  and few days). In fact, the gyration time of the PeV particles is of order the duration of the flare themselves. Hence, the particles must be accelerated over a sub-Larmor timescale, i.e., the acceleration process must be extremely efficient. Diffuse shock-acceleration is not adequate to explain the flares because it operates only over multiple gyrations of the particles moving back and forth through the shock front. In addition, the inferred flaring particle spectrum can be as hard as $d{\rm N}/d\gamma\propto \gamma^{-1.6}$ \citep{2012ApJ...749...26B} which is inconsistent with the steep power-law (i.e., of index $\lesssim -2$) usually expected in diffuse shock-acceleration (e.g., \citealt{1987PhR...154....1B, 2009ApJ...698.1523S}).

\subsection{Multiwavelength follow-up of Crab PWN flares}

During the gamma-ray flares the Crab was observed in radio, infrared, optical, X-ray and TeV energies, but no substantial variation in the flux emission at these wavelengths was measured. After the discovery of the gamma-ray flares, the Chandra X-ray observatory started to observe the Crab approximately every month. Five observations were carried out during the major gamma-ray flare of April 2011. The bright Anvil region (see Figure \ref{linefree}) and several other regions (that are known to be active) exhibit time variability during the flaring activity time. Nevertheless, despite these hints in the X-ray data,  there is no evidence for statistical significant variations associated with the flare \cite{2013ApJ...765...56W}.
The near-IR observations performed by Keck's NIRC2 revealed that the inner knot (knot-1) was slightly brighter when compared to previous observations. Indeed knot-1, which is the brightest feature from the Nebula in the near infrared energy band, was reported to show flux variation at this wavelength of the order of $\simeq  35\%$. However this variation is well within the range typically observed from this region. The radio observations performed with VLA did not reveal anything interesting. No other point source, a part from the pulsar, was found. Therefore, no ``smoking gun'' has been identified from the X-ray, near-IR and radio observations.

In terms of X-ray and optical counterparts of the Crab flares, we can describe two scenarios:

\begin{enumerate}
  \item A simultaneous brightening of X-rays and optical locations associated with the gamma-ray flares. This scenario would favor ``shock-driven'' power-law models of particle acceleration;
  \item A delayed response of the optical/X-ray emission, with a timescale that depends on the radiative cooling properties of the accelerated particle population. This scenario would favor an extremely efficient acceleration mechanism, likely to saturate the particle energy to a maximum value, and that could be modeled with a quasi-monoenergetic particle distribution;
\end{enumerate}

The absence of a strongly enhanced X-ray and/or optical location in the Nebula in coincidence with the gamma-ray flares tends to exclude the first scenario.
However in \cite{2013ApJ...765...56W} the possibility to model the photon spectrum of the flare of April 2011 with a power-law with index $\Gamma \sim 1.3$ connecting gamma rays and X rays was investigated and could not be ruled out.

Because of the poor angular resolution of gamma-ray telescopes, localizing the emission site of the Crab flares is a big challenge. Several regions can be considered as candidate for the acceleration and emitting region. Among them, we can identify 3 particularly interesting possibilities:

\begin{enumerate}
  \item instabilities in the Anvil at the South-East jet base. Variability in this region
   was detected by optical and X-ray observations both during the September 2010 and April 2011 gamma-ray flares \citep{2011Sci...331..736T,2013ApJ...765...56W}. MHD simulations of the Crab South-East jet \citep{2013MNRAS.436.1102M} revealed substantial jet deviation and magnetic dissipation. The jet, fed by highly magnetized and relativistic plasma \citep[$\sigma \sim 1\--10$,][]{2013MNRAS.436.1102M}, could be a region of magnetic reconnection. Kink instability in the jet could trigger magnetic reconnection and consequent particle acceleration.
  \item tearing mode and reconnection on the termination shock ``ring''.
  This region is known to be highly variable. The three most variable spots during the April 2011 flare are located along the ring. The highly variable wisps are observed to originate from this region. Recent simulations \citep{2014MNRAS.438..278P} recently proposed that efficient magnetic reconnection could take place in the Nebula right after the termination shock;
  \item variation in the observed emission from knot-1. Variability up to $\sim 20\--30\%$ is known. Its variability is interpreted as a variable Doppler factor $\delta$, but no indication for optical variability has been observed so far. Simulations \citep{2009MNRAS.400.1241C, Komissarov2011} show that at 100 MeV the inner knot is the brightest feature of the Nebula, and that the magnetic field can be up to 10 times larger than the average Nebular magnetic field.
\end{enumerate}

\begin{figure}[]
\centering
\includegraphics[width=12cm]{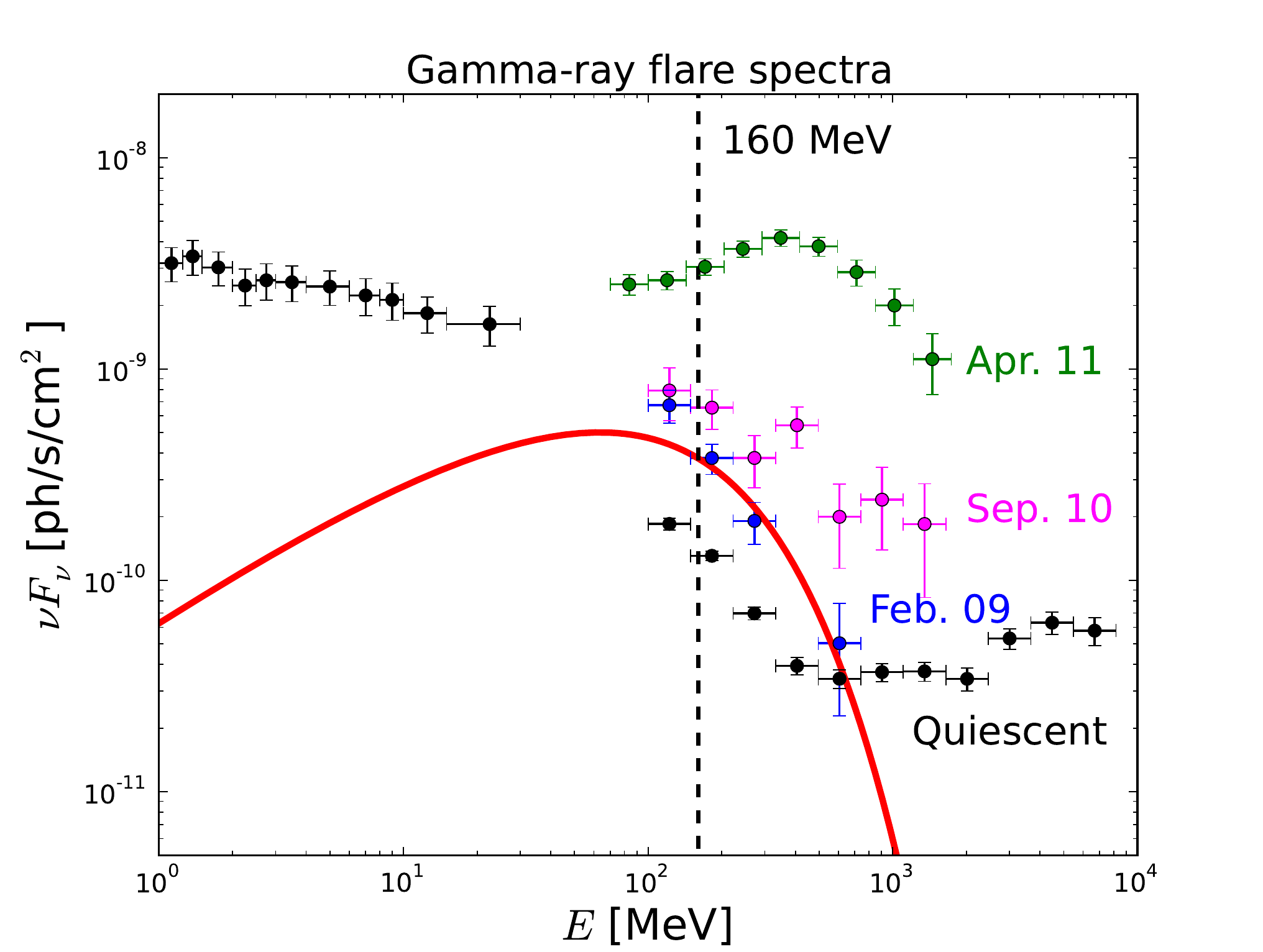}
\caption{Measured gamma-ray spectra of the Crab Nebula in the 1~MeV-10~GeV band. Black data-points show the ``quiescent'' spectra of the Nebula. The green, magenta, and blue points are respectively the $>100$ MeV spectra during the April 2011, September 2010 and February 2009 flares measured by the {\em Fermi}-LAT \citep{2014RPPh...77f6901B}. The red solid line is the spectra obtained with 3D PIC simulations with radiation reaction of a reconnecting current sheet in the Crab Nebula \citep{2014ApJ...782..104C}. The model assumes that the Nebula is 2~kpc away.}
\label{fig_spectra}
\end{figure}

\subsection{Proposed models of the flares}\label{sect_model}

It is usually expected that synchrotron spectrum observed in astrophysical sources cuts off far below 100~MeV \citep{1970RvMP...42..237B, 1983MNRAS.205..593G}. This limit is given by the balance of two antagonist forces acting on the particles: (i) the accelerating electric force, and (ii) the radiation-reaction-force opposite to the particle's motion due to the emission of synchrotron photons. Hence, there is a maximum energy limited by synchrotron losses rather than by the size of the accelerator (e.g., \citealt{2002PhRvD..66b3005A, 2003PhRvE..67d5401M}). Then, one can show that the maximum (critical) synchrotron photon energy should be
\begin{equation}
\epsilon^{\rm sync}_{\rm max}=\frac{9m_{\rm e}c^2}{4\alpha_{\rm F}}\left(\frac{E}{B_{\perp}}\right)\approx 160\times\left(\frac{E}{B_{\perp}}\right)~{\rm MeV},
\end{equation}
where $m_{\rm e}$ is the rest mass of the electron, $\alpha_{\rm F}\approx 1/137$ is the fine structure constant, $E$ the electric field, and $B_{\perp}$ the magnetic field perpendicular to the particle's direction of motion. A particle accelerated above the radiation reaction limit would radiate away most of its energy within a sub-Larmor cycle. In most cases, ideal MHD applies, $E\ll B$, so the synchrotron spectrum should cut $\ll 100~$MeV. This rule of thumb generally applies well to astrophysical sources, and in particular to the quiescent Crab Nebula where the synchrotron spectrum turns over at about the 100~MeV limit \citep{1996ApJ...457..253D, 2010ApJ...708.1254A}. However, the flaring emission is systematically extending significantly above 160~MeV, up to about 1~GeV (see Fig.~\ref{fig_spectra}). Unless the emitting region moves at highly relativistic speeds ($\gtrsim 0.9~c$), it implies that $E\gtrsim 5 B_{\perp}$, suggesting that a non-ideal MHD process may be at work. This is, once again, difficult to explain with classical models of particle acceleration.

All of the estimates derived above from observations are quite conservative, in a sense that we ignored the effect of beaming (geometrical or relativistic), spatial and/or temporal inhomogeneities. Every model proposed so far is taking advantage of one or more of these effects to alleviate the tight constraints imposed by the flares. For instance, one feature commonly invoked in models is a strong inhomogeneity of the flaring region, in particular in the magnetic field structure. \citet{2012MNRAS.421L..67B} proposed that the flares occur around the equatorial belt where the annihilation of the striped pulsar wind at the shock accelerates particles \citep{2003MNRAS.345..153L, 2007A&A...473..683P, 2011ApJ...741...39S} and, crucially for this model, generates magnetic turbulence. They find that a concentration of fluctuating magnetic field can generate an intermittent, strongly polarized gamma-ray signal that is most pronounced at the high-energy end of the synchrotron spectrum. In this model, synchrotron photons above $100~$MeV can be emitted if the magnetic field varies over a timescale shorter than the synchrotron cooling time of the particles which is determined only on the RMS value of the field. The observed gamma-ray variability would then be given by the statistical properties of the magnetic fluctuations (lifetime, amplitude).

The model by \citet{2013ApJ...763..131T} also relies on a highly inhomogeneous turbulent flow, but in which the coherence length of the magnetic field, $\lambda_B$, is extremely short compared with the cooling length of the particles and even compared with the formation length of the synchrotron photons, $\lambda_{\rm sync}=m_{\rm e}c^2/eB\gg \lambda_B$, \footnote{This might be challenging to achieve in relativistic shocks \citep{2009ApJ...707L..92S}.}. In this regime, the particles emit the so-called ``jitter'' radiation rather than the classical synchrotron radiation (see, e.g., \citealt{2000ApJ...540..704M}). The cooling rate of the particles is identical to synchrotron but the emitted spectra can differ significantly. In particular, the critical photon energy increases by a factor $\epsilon_{\rm jitt}/\epsilon_{\rm sync}\sim\lambda_{\rm sync}/\lambda_B\gg 1$ in the jitter regime, and hence $>100$~MeV gamma rays could be emitted even by particles below the radiation reaction limit. In addition, jitter radiation can produce a harder spectrum than synchrotron, specifically, $F_{\nu}\propto\nu$ instead of $F_{\nu}\propto \nu^{1/3}$ for a mono-energetic population of particles. However, we note that jitter radiation is not needed to explain the flare spectra. So far, observations are fully consistent with synchrotron radiation \citep{2012ApJ...749...26B, 2013ApJ...765...56W}, but this could be tested in the future.

A more natural way to explain the Crab flares is to invoke a relativistic bulk motion of the flaring region with a modest Lorentz factor $\Gamma\gtrsim 2$. Indeed, the relativistic motion of the source can boost $\lesssim 100~$MeV synchrotron photons emitted in the co-moving frame by a $\sim\Gamma$ factor above the radiation reaction limit \citep{2011ApJ...730L..15Y, 2011MNRAS.414.2229B, 2011MNRAS.414.2017K, 2012MNRAS.422.3118L, 2012MNRAS.426.1374C}. A Doppler boost would also relax the tight constraints on the size of the region and the duration of the flare in the co-moving frame, and beam the emission in the frame of the observer which would also help at reducing the energetic constraints\footnote{This solution is often proposed to account for the ultra-rapid gamma-ray flares in blazars.}. Lastly, the Doppler beaming would explain the observed correlation between the gamma-ray flux and the cut-off energy in the spectra during the April 2011 flare \citep{2012ApJ...749...26B}. Although this simple solution solves many problems at once, there is still no definite evidence that such a relativistic flow exists in the Crab Nebula. Observations show only mildly relativistic flows with proper motion of order half the speed of light (e.g., \citealt{2008ARAandA..46..127H}). However, in principle, highly relativistic flows could emerge in the polar regions of the Nebula because the relativistic shock is oblique and magnetized (weak shock) at high latitudes (e.g., \citealt{2012MNRAS.427.1497L, 2013MNRAS.428.2459K}). In particular, \citet{2011MNRAS.414.2017K} argued that the flares may originate from the emission from the oblique shock Doppler-boosted towards the observer, that they associate with the well-known bright compact structure near the pulsar (the so-called ``inner knot''). Unfortunately, the brightness of the knot does not show any variations correlated with the gamma-ray flares (see \citealt{2011Sci...331..736T, 2011A&A...533A..10L, 2013ApJ...765...56W} and references therein) contrary to what the model predicts.

Alternatively, \citet{2011ApJ...737L..40U} proposed that magnetic reconnection could accelerate particles well above the radiation reaction limit, and hence result in the emission of $>100~$MeV synchrotron radiation. Indeed, as pointed out by \citet{2004PhRvL..92r1101K}, inside a reconnection layer the magnetic field is small and even vanishes at its center while the reconnection electric field is maximum, i.e., we are in the situation where $E\gg B$. Thus, in principle, a particle trapped deep inside the reconnection layer could be linearly accelerated by the electric field to arbitrary high energies with little synchrotron losses. The maximum energy of the particle would be limited only by the length of the reconnection layer $L$, i.e., $\mathcal{E}_{\rm max}\sim e E L=e\beta_{\rm rec}B_0 L$ where $\beta_{\rm rec}=E/B_0$ is the dimensionless reconnection rate, $E$ is the reconnection electric field and $B_0$ is the reconnecting magnetic field. Using a test-particle approach with prescribed static fields, \citet{2007A&A...472..219C} and \citet{2012ApJ...746..148C} showed that the high-energy particles are naturally trapped and confined deeply inside the layer, where they follow the relativistic analog of Speiser orbits \citep{1965JGR....70.4219S, 2011ApJ...737L..40U}. This scenario was successfully tested using state-of-the-art particle-in-cell (PIC) simulations of 2D and 3D relativistic reconnection with guide field and, more importantly, with the radiation reaction force \citep{2013ApJ...770..147C, 2014ApJ...782..104C}. Furthermore, these studies revealed that a natural outcome of relativistic reconnection is the strong beaming and bunching of the energetic particles (see also \citealt{2012ApJ...754L..33C}). The combination of both effects results in several bright ultra-rapid synchrotron flares above $100$~MeV consistent with the observed intra-flare variability ($\lesssim 6$~hrs) if, by chance, the beam crosses the line of sight of a distant observer. Thanks to beaming, this model can also explain the overall energetics of the flares (Fig.~\ref{fig_spectra}), the flux/cut-off energy correlation and, at least qualitatively, the hard particle spectrum\footnote{Recent studies show that reconnection produces hard particle spectra $d{\rm N}/d\gamma\propto \gamma^{-1},~\gamma^{-1.5}$ for $\sigma\gg1$ \citep{2014ApJ...783L..21S, 2014PhRvL.113o5005G, 2014arXiv1409.8262W}.}.

The reconnection scenario works best in a highly magnetized flow (i.e., $\sigma\gg 1$) which may be hard to find in the nebula, except in the polar regions and in the jets. As predicted by \citet{1998ApJ...493..291B} and as recently shown by \citet{2011ApJ...728...90M, 2014MNRAS.438..278P, 2013MNRAS.436.1102M}, the polar regions and the jets are subject to kink instabilities which results in important magnetic dissipation (see Sect.~\ref{sect_sigmapb}), and may ultimately power the Crab flares \citep{2012ApJ...746..148C, 2012MNRAS.427.1497L, 2013MNRAS.428.2459K, 2013MNRAS.436.1102M}.

\subsection{Comparison with GeV flares in the PSR~B1259--63/LS~2883  binary}\label{sect_model}

The gamma-ray binary PSR~B1259$-$63 that contains a 48-ms pulsar in a
3.4-yrs eccentric orbit around an O star (see Sect.~1.4), is a well-known TeV gamma-ray emitter
\citep{2005A&A...442....1A, 2013A&A...551A..94H}. The very-high energy
radiation is often interpreted as inverse Compton scattering of the UV
stellar photons on relativistic pairs accelerated near the shock front
between the pulsar wind and the stellar wind (e.g.,
\citealt{1997ApJ...477..439T, 1999APh....10...31K,
2007MNRAS.380..320K}). The model predicts a maximum of GeV gamma-ray
emission close to periastron which was indeed observed for the first
time by the Fermi-LAT during the 2010 periastron passage
\citep{2011ApJ...736L..10T, 2011ApJ...736L..11A}. A few weeks after
the peak of the periastron emission faded away, and against all
expectations, a bright flare appeared in the Fermi data. The flare is
about 10-times brighter than the predicted emission at periastron,
which represents a gamma-ray luminosity comparable to the pulsar
spin-down power. A radiative efficiency close to 100\% is in principle
achievable with inverse Compton scattering, but the density of stellar
photons is far too low to explain the flux at these phases, unless
there are extra sources of radiation close to the pulsar (see, e.g.,
\citealt{2012ApJ...752L..17K, 2013A&A...557A.127D}), or significant
Doppler boosting of the emission towards the observer
\citep{2010A&A...516A..18D, 2012ApJ...753..127K}. The flare peaks at
300 MeV and is seen only in the GeV band, which suggests that the
particle energy distribution must be very narrow. These properties
remind us of the Crab-flare events (see Sect.~\ref{sect_puzzle}).
Although both flares share similar properties, they have also
important differences. The week-long flares in the Crab reach at most
1\% of the Crab pulsar spin-down power, whereas the highest day-average flux reaches nearly 100\% of the pulsar
spin-down flux during flaring period associated with the second disk
passage. Unlike Crab PWN the flares in B1259--63 appear to be periodic. i.e.
they occurred at similar binary phases (close to the second disk passage) during the past two binary cycles
\citep{2015ApJ...798L..26T, 2015arXiv150406343C}.

While it may be that the underlying nature of  flares in both
cases is the synchrotron radiation associated with  reconnection in
the magnetized relativistic plasma, the details should differ.  In the
Crab  PWN, the reconnection can be driven by the growth of
instabilities or other random process, while in B1259--63, it can be
driven  by the magnetic field distortion and compression caused by the
pulsar   passage through the excretion disk.  It is possible that the reconnection 
happens in the tail of the PWN after the pulsar passage through the disk which could explain the 
delay between the GeV flare and the peak of X-ray flux \citep{2015ApJ...798L..26T, 2015arXiv150406343C}. An alternative
scenario, considered by \cite{2012ApJ...752L..17K}, where the flare is
due to the IC radiation associated with the increase in volume
occupied by the unshocked pulsar wind when the excretion disk is
strongly perturbed, also remains a possibility. One may be able to
differentiate between the two scenarios once the $\gamma$-ray
variability timescales and the state of the disk during the pulsar
passage are better probed by the observations.

\begin{acknowledgements}
We thank George Pavlov for his valuable comments and  discussions. We are grateful to  Blagoy Rangelov for creating the merged F550M Crab image from the archival {\sl HST} data and help with the manuscript editing. We acknowledge support from NASA grants GO3-14084X, GO3-14057C, G03-14082A, NNX09AC81G,  NNX09AC84G, HST-GO-13043.09, and G02-13085C. BC acknowledges support from the Lyman Spitzer Jr. Fellowship awarded by the Department of Astrophysical Sciences at Princeton University, and the Max-Planck/Princeton Center for Plasma Physics. YL acknowledges support from Israeli Science Foundation under the grant 719/14.
\end{acknowledgements}

\bibliographystyle{aps-nameyear}      
\bibliography{biblio_total_final_13Feb}                

\end{document}